\documentclass[12pt]{article}
\usepackage[bookmarks,bookmarksnumbered]{hyperref}
\usepackage{amsmath,XoohmE}
\usepackage{graphicx,color}
\usepackage{booktabs,multirow}

\def\ha{{\hat\alpha}}
\def\hb{{\hat\beta}}
\def\hg{{\hat\gamma}}
\def\hd{{\hat\delta}}
\def\hK{{\hat{\rm K}}}
\def\hL{{\hat{\rm L}}}
\def\ttmes{\mathbin{\rotatebox{45}{\!{\tt\#}}}\!}
\def\Bf{\sf\bfseries}
\def\he{\hat{\rm e}}
\def\rD{{\rm D}}
\def\Gf{G_\text{eff}}
\def\rI{{\rm I}}
\def\img{\mathop\text{\rm img}\nolimits}
\def\rJ{{\rm J}}
\def\rK{{\rm K}}
\def\tl{{\tilde\ell}}
\def\rL{{\rm L}}
\def\rR{{\rm R}}

\def\ddt{\partial_\tau}
\def\hi{{\hat\imath}}
\def\hj{{\hat\jmath}}
\def\vC#1{\vcenter{\hbox{\hss#1\hss}}}
\def\bs#1{\boldsymbol{#1}}
\def\sO{\mathop\textsl{O}}
\def\Pin{\mathop\textsl{Pin}}

\def\SO{\mathop\textsl{SO}}
\def\Span{\mathop\textsl{Span}}
\def\Spin{\mathop\textsl{Spin}}
\def\SZ{\mathop\textsl{Spin}(4){\times}\ZZ_2}
\def\spin{\mathop\mathfrak{spin}}
\def\Sp{\mathop\textsl{Sp}}	
\def\SU{\mathop\textsl{SU}}

\def\Sym{\mathop\text{\rm Sym}\nolimits}

\def\sU{\mathop\textsl{U}}
\def\wt{\mathop\textrm{wt}}	
\definecolor{WIP}{rgb}{.8,0,.2}
\def\1{\raisebox{-2pt}{}}
\def\<#1|#2>{\leavevmode\setbox7=\hbox{\colorbox{yellow}{#2}}\copy7\kern-\wd7%
              \smash{\hbox to0pt{\color{red}\vrule height2.5ex width.5pt
               \vrule height9pt depth-8.5pt width15pt\kern-14pt
                \raisebox{10pt}{\footnotesize\it\bfseries#1}}}\kern\wd7}

\def\be{\begin{equation}}
\def\ee{\end{equation}}
\def\fracm#1#2{\hbox{\large{${\frac{{#1}}{{#2}}}$}}}

\definecolor{Green}  {rgb}{0.10,0.61,0.22}
\definecolor{Orange} {rgb}{1.00,0.50,0.15}
\definecolor{Red}    {rgb}{0.78,0.00,0.12}
\definecolor{Purple} {rgb}{0.42,0.15,0.45}
\definecolor{Turque} {rgb}{0.00,0.65,0.85}
\definecolor{Blue}   {rgb}{0.00,0.00,1.00}
\definecolor{Magenta}{rgb}{1.00,0.00,1.00}
\definecolor{Gold}   {rgb}{1.00,0.85,0.35} 
\definecolor{Seaweed}{rgb}{0.01,0.24,0.09}
\definecolor{Brown}  {rgb}{0.43,0.26,0.32}
\definecolor{Grey}  {rgb}{0.40,0.40,0.40}
\def\C#1#2{{\ifcase#1\or
             \color{Green}\or \color{Orange}\or \color{Red}\or
              \color{Purple}\or \color{Turque}\or \color{Blue}\or
               \color{Magenta}\or \color{Gold}\or \color{Seaweed}\or
                \color{Brown}\else\color{Grey}\fi#2}}

\def\Ft#1{\,\footnote{#1}}
\newdimen\parshift\parshift=\parindent
\catcode`@=11
 \long\def\@footnotetext#1{\insert\footins{\reset@font\footnotesize\interlinepenalty%
  \interfootnotelinepenalty\splittopskip\footnotesep\splitmaxdepth\dp\strutbox%
   \floatingpenalty\@MM\hsize\columnwidth\addtolength{\hsize}{-2\parindent}
    \@parboxrestore\protected@edef\@currentlabel{\csname p@footnote\endcsname\@thefnmark}
      \color@begingroup
       \@makefntext{\rule\z@\footnotesep\baselineskip=11pt\ignorespaces#1\@finalstrut\strutbox}
        \color@endgroup}}
 \long\def\@makefntext#1{\hglue\parshift
                         \vbox{\noindent\hb@xt@0em{\hss\@makefnmark}#1}}
\catcode`@=12
 \font\rOpe=cmsy10                        
 \def\ktl{{\hbox{\rOpe\char'170}}}        
 \def\kbl{{\hbox{\rOpe\char'170}}}        
 \def\kcr{{\reflectbox{\rOpe\char'170}}}        
 \def\ktr{{\reflectbox{\rOpe\char'170}}}        
 \def\kbr{{\reflectbox{\rOpe\char'170}}}        
 \def\Border{\vbox{\hsize0pt
        \setlength{\unitlength}{1mm}
        \newcount\xco
        \newcount\yco
        \xco=-21
        \yco=12
        \begin{picture}(0,0)(-7.5,0)
        \put(\xco,\yco){$\ktl$}
        \advance\yco by-1
        {\loop
        \put(\xco,\yco){$\kcr$}
        \advance\yco by-2
        \ifnum\yco>-240
        \repeat
        \put(\xco,\yco){$\kbl$}}
        \xco=170
        \yco=12
        \put(\xco,\yco){$\ktr$}
        \advance\yco by-1
        {\loop
        \put(\xco,\yco){$\kcr$}
        \advance\yco by-2
        \ifnum\yco>-240
        \repeat
        \put(\xco,\yco){$\kbr$}}
        \put(-19.5,13){\scalebox{.54}{State University of New York
            Physics Department|University of Maryland Center for
            String and Particle  Theory \&\ Physics Department|%
            Howard University Physics \&\ Astronomy Department}}
        \put(-19.5,-241.5){\scalebox{.649}{University of Alberta Mathematical
            and Statistical Sciences Department|Pepperdine University Natural
            Sciences Division|Bard College Mathematics Program}}
        \end{picture}
        \par\vskip-8mm}}
\definecolor{UMred}{rgb}{.9,.05,.2}
 \def\UMbanner{\vbox{\hsize0pt
        \setlength{\unitlength}{.4mm}
        \thicklines
        \begin{picture}(0,0)(-30,-10)
        \put(165,16){\line(1,0){4}}
        \put(170,16){\line(1,0){4}}
        \put(180,16){\line(1,0){4}}
        \put(175,0){\line(1,0){4}}
        \put(180,0){\line(1,0){4}}
        \put(185,0){\line(1,0){4}}
        \put(169,0){\line(0,1){16}}
        \put(170,0){\line(0,1){16}}
        \put(179,0){\line(0,1){16}}
        \put(180,0){\line(0,1){16}}
        \put(184,0){\line(0,1){16}}
        \put(185,0){\line(0,1){16}}
        \put(169,16){\oval(8,32)[bl]}
        \put(170,16){\oval(8,32)[br]}
        \put(179,0){\oval(8,32)[tl]}
        \put(185,0){\oval(8,32)[tr]}
        \end{picture}
        \par\vskip-6.5mm
        \thicklines}}

 \SfTitles 
 \allowdisplaybreaks
 \seceq 

\begin{document}
\thispagestyle{empty}
\vbox{\Border\UMbanner}
 \noindent
 \today;~\hourmin
  \hfill\smash{\parbox[t]{50mm}{\raggedleft SUNY-O/673\\[-1mm]
                                            UMDEPP 09-032}}
  \vfill
 \begin{center}
{\LARGE\sf\bfseries\boldmath
  Effective Symmetries of the Minimal Supermultiplet\\[2mm]
  of the $N\,{=}\,8$ Extended Worldline Supersymmetry
 }\\*[5mm]
{\sf\bfseries M.G.\,Faux$^a$,
              S.J.\,Gates, Jr.$^b$
              and
              T.\,H\"{u}bsch$^c$}\\*[2mm]
{\small\it
  $^a$Department of Physics,\\[-1mm]
      State University of New York, Oneonta, NY 13825%
  \\[-4pt] {\tt  fauxmg@oneonta.edu}
  \\
  $^b$Center for String and Particle Theory,\\[-1mm]
      Department of Physics, University of Maryland, College Park, MD 20472%
  \\[-4pt] {\tt  gatess@wam.umd.edu}
  \\
  $^c$Department of Physics \&\ Astronomy,\\[-1mm]
      Howard University, Washington, DC 20059
  \\[-4pt] {\tt  thubsch@howard.edu}
 }\\[5mm]
  \vfill
{\sf\bfseries ABSTRACT}\\[3mm]
\parbox{144mm}{\parindent=2pc\noindent
A minimal representation of the $N\,{=}\,8$ extended worldline supersymmetry, known as the {\em\/ultra-multiplet\/}, is closely related to a family of supermultiplets with the same, $E_8$ chromotopology. We catalogue their effective symmetries and find a $\SZ$ subgroup common to them all, which explains the particular basis used in the original construction.
We specify a constrained superfield representation of the supermultiplets in the ultra-multiplet family, and show that such a superfield representation in fact exists for all adinkraic supermultiplets. We also exhibit the correspondences between these supermultiplets, their Adinkras and the $E_8$ root lattice bases. Finally, we construct quadratic Lagrangians that provide the standard kinetic terms and afford a mixing of an even number of such supermultiplets controlled by a coupling to an external 2-form of fluxes.
 }
\end{center}
 \vfill

\clearpage\setcounter{page}{1}
\section{Introduction}
 \label{s:I}
An off-shell model in one bosonic dimension\footnote{A two-dimensional on-shell model under a compactification produces a one-dimensional off-shell model.} identifiable as the worldline, with $N=8$ supersymmetry was named {\em\/ultra-multiplet\/} in its inaugural presentation in the physics literature\cite{rGR0}; see also Refs.\cite{rBIKL03,rBIKL04,rILS,rGRT}.  The ultra-multiplet was introduced in a manifestly $\SZ$-symmetric notation; see below and in particular Appendix~\ref{a:Spin}. By means of a systematic component field redefinition by now known as {\em\/node-raising/lowering\/}\cite{rGR2,rA,r6-1}, this supermultiplet is seen to be closely related to a {\em\/family\/} of supermultiplets, identifiable with the ``root superfield'' of Ref.\cite{rGLP,rPT,rBKMO}. All members of this family have equivalent chromotopologies\cite{r6-1,r6-3} and are describable in the common $\SZ^e$-basis.
 However, as we show herein, these supermultiplets have different effective symmetries\Ft{Unlike {\em\/dynamical\/} symmetries, which are determined by the action functional, these symmetries characterize the supermultiplets themselves and so also every model built from them.}. The concept of such symmetries is ubiquitous in fundamental physics, and we trust the general motivation for their study is self-evident. In addition, this is the lowest $N$-extended supersymmetry where the minimal supermultiplet is a maximal, $(\ZZ_2)^{N/2}$-quotient, and is unique in having eight bosons and eight fermions, transformed into each other by eight $Q$'s|a new {\em\/eightfold way\/} paradigm.
 We also couple the supermultiplets in the ultra-multiplet family to background fluxes, exhibiting a super-Zeemann effect and providing another example of the general framework of Ref.\cite{r6-7a}.

Worldline $N=8$ supersymmetry is generated by eight supercharges $Q_\rI$ and the worldline Hamiltonian, $H=i\ddt$, satisfying the relations
\begin{subequations}
 \label{eSuSy}
 \begin{equation}
  \big\{\,Q_\rI\,,\,Q_\rJ\,\big\} = 2\,\d_{\rI\rJ}\,H,\qquad
  \big[\,H\,,\,Q_\rI\,\big] = 0,\qquad
  (Q_\rI)^\dag=Q_\rI,\qquad H^\dag=H.
 \label{eSuSyQ}
\end{equation}
In superspace, there also exist eight super-derivatives $\rD_\rI$, satisfying
\begin{equation}
  \big\{\,\rD_\rI\,,\,\rD_\rJ\,\big\} = 2\,\d_{\rI\rJ}\,H,\qquad
  \big[\,H\,,\,\rD_\rI\,\big] = 0 = \big\{\,Q_\rI\,,\,\rD_\rJ\,\big\}.
 \label{eSuSyD}
\end{equation}
\end{subequations}
Here, $\rI,\rJ=1,\cdots,8$, and the explicit appearance of $\d_{\rI\rJ}$ as an invariant symbol implies that the maximal symmetry of the system\eq{eSuSy} is $\textsl{O}(8)$, of which the $Q$'s span the 8-vector representation, as do independently the $\rD$'s. In the study of representations of the algebra\eq{eSuSy}, we will often be able to omit the $(Q_\rI,\rD_\rI)\to(-Q_\rI,-\rD_\rI)$ reflection operations for any fixed $\rI\in\{1,\cdots,8\}$, thus reducing to the $\SO(8)$ subgroup. In turn however, we will have to pass to the double-cover $\Spin(8)$, and also find use for its $\ZZ_2$-extension, $\Pin(8)$, although less frequently as a symmetry.

In Section~\ref{s:UM}, we summarize the basic facts about the ultra-multiplet, and  then re-analyze it in Section~\ref{s:E8} in terms of Adinkras\cite{rA,r6-1,r6-3,r6-3.2,r6-3.4}.
 This leads us to catalogue, in Section~\ref{s:SymUM}, the effective symmetries of the ultra-multiplet and its node-raised relatives. This reveals, in Section~\ref{s:Roots}, the group-theoretic reason behind the existence of the basis that is computationally effective throughout the entire family of supermultiplets\cite{rGR0} and a triality-rotation thereof.
 Returning to more physical applications, supermultiplets from the ultra-multiplet family are coupled to external (background) fluxes in Section~\ref{s:SZEM}, and Section~\ref{s:CO} collects our concluding comments and outlook.
 Appendix~\ref{a:D2Q} relates the {\em\/adinkraic\/} methods to the more traditional supersymmetry techniques\cite{r1001,rPW,rWB,rBK};
 Appendix~\ref{a:Same} details the various notions of isomorphism between supermultiplets, and so specifies what we mean by ``the ultra-multiplet;''
 Appendix~\ref{a:Spin} clarifies the general nature of the $\Spin(N)\times\ZZ_2^e\subset\Spin(2N)$ basis;
 Appendix~\ref{a:E8} presents the correspondences between the $E_8$ algebra, root lattice bases, the $e_8$ binary code, the Adinkras depicting the ultra-multiplet family and this supermultiplet family itself.

\section{Field Theory of the Free Ultra-Multiplet and Family}
 \label{s:UM}
A translation of the notation of Ref.\cite{rGR0} informs us that the component fields of the ultra-multiplet may be identified as eight real bosons: two scalars, $A$, $B$, and two rank-2 antisymmetric tensors, $A_{\sss\ha\hb}$, and $B_{\sss\ha\hb}$ of Spin(4); the indices $\ha$, $\hb$, {\em\/etc\/}. take on values $1,\cdots,4$.  The Levi-Civita tensor $\ve_{\sss\ha\hb\hg\hd}$ involving these indices can be used to impose a self-duality condition on the rank-2 antisymmetric tensors $A_{\sss\ha\hb}$, and $B_{\sss\ha\hb}$ according to
\begin{equation}
 A_{\sss\ha\hb} ~=~ +\,\fracm12\,
  \ve_{\sss\ha\hb}{}^{\sss\hg\hd}\,A_{\sss\hg\hd},\qquad
 B_{\sss\ha\hb} ~=~ -\,\fracm12\,
  \ve_{\sss\ha\hb}{}^{\sss\hg\hd}\,B_{\sss\hg\hd}.
\label{UM1}
\end{equation}
As representations of $\Spin(4)=\SU(2)\times\SU(2)$, we identify $A,B\sim({\bf1,1})$, $A_{\sss\ha\hb}\sim({\bf3,1})$ and $B_{\sss\ha\hb}\sim({\bf1,3})$.
 The supermultiplet also includes eight fermions $\psi\1_{\sss\hK\,\ha}\sim({\bf2,2})_\pm$. As indicated, they carry a $\hK$-type index, which takes on values of ``$+$'' and ``$-$.'' The values of such indices are additive and stem from a $\Spin(2)=\sU(1)$ charge. However, the explicit use of the diagonal matrices $\d_{\sss\hK\hL}$ and $(\s^3)_{\sss\hK\hL}$ as invariants in \Eq{UM2.3} along with the off-diagonal $(\s^1)_{\sss\hK\hL}$ and $\ve_{\sss\hK\hL}$
 indicates covariance only with respect to a discrete subgroup the net total value of these indices is to be taken (mod~2). This then distinguishes only even {\em\/vs.\/} odd numbers of $\hat{\rK}$-type indices, \ie, tensors {\em\/vs.\/} spinors, and so identifies these indices as $\Spin(1)=\ZZ_2$ labels. In addition however, the specific basis\eq{UM2} does consistently distinguish $\hK=+1$ from $\hK=-1$, and so the conjugate spinors of a $\Spin(2)\supset\Spin(1)$. We therefore ``extend'' $\Spin(1)=\ZZ_2$ into ``$\ZZ_2^e$,'' but hasten to emphasize that the {\em\/group structure\/} denoted as ``$\SZ^e$'' only includes this $\ZZ_2=\Spin(1)$, not its ``extension''\Ft{This $\ZZ_2=\Spin(1)$ assigns an additive, (mod~2) charge ``0'' to tensors and ``1'' to spinors. Augmenting this by distinguishing $\pm1$ as corresponding to the two different spinors of $\Spin(2)=U(1)\supset\Spin(1)=\ZZ_2$|but retaining the (mod~2) structure otherwise|leads to an algebraic structure that lacks associativity, and so does not form a proper symmetry group. By ``$\ZZ_2^e$ {\em\/group\/}'' we then always mean just this $\Spin(1)=\ZZ_2$.}; see Appendix~\ref{a:Spin}.

For future convenience, we introduce for each component field an {\em\/a priori\/} unconstrained and unprojected superfield of the same name (but set in bold font), so that the lowest component of each such superfield is the said component field. To wit,
\begin{equation}
  A={\bf A}|,\quad
  A_{\sss\ha\hb}={\bf A}_{\sss{\ha\hb}}|,\quad
  B={\bf B}|,\quad
  B_{\sss\ha\hb}={\bf B}_{\sss{\ha\hb}}|,\quad
  \j\1_{\sss\hK\,\ha}=\bs{\J}\1_{\sss\hK\,\ha}|,
 \label{UM1.5}
\end{equation}
where right-delimiting ``$|$'' denotes evaluation at $\q^{\sss\hK\,\ha}\to0$ in superspace.
 The supersymmetry transformations between the component fields $(A,B,A_{\sss\ha\hb},B_{\sss\ha\hb}|\j\1_{\sss\hK\ha})$ may then be summarized by 
a set of super-differential equations relating the corresponding superfields; for details, see Appendix~\ref{a:D2Q}.

To describe a Valise\Ft{The term ``Valise,'' in the language developed to describe the associated Adinkras\cite{rGR2,r6-1,r6-3,r6-3.2,r6-3.4}, indicates that all the bosons possess the same engineering units, and similarly all the fermions, offset by $\inv2$: $[\j_\hj]=[\f_i]\pm\inv2$.} supermultiplet, these 
super-differential relations are
\begin{subequations}
  \label{UM2}
\begin{alignat}{3}
 \rD_{\sss{\hK\,\ha}}{\bf A}
  &= i\,(\s^3)_{\sss{\hK}}{}^{\sss{\hL}}\,\bs{\J}_{\sss{\hL\,\ha}},
 \quad&\quad
 \rD_{\sss{\hK\,\hg}}{\bf A}_{\sss{\ha\hb}}
  &= i\,\big[\,\d_{\sss{\hg[\ha}}\,\bs{\J}_{\sss{\hK\,\hb]}}
         ~+~\ve\1_{\sss{\ha\hb\hg}}{}^{\sss\hd}\,
             \bs{\J}\1_{\sss{\hK\,\hd}}\,\big],
 \label{UM2.1}\\[2mm]
 \rD_{\sss{\hK\,\ha}}{\bf B}
  &= i\,(\s^1)_{\sss{\hK}}{}^{\sss{\hL}}\,\bs{\J}_{\sss{\hL\,\ha}},
 \quad&\quad
 \rD_{\sss{\hK\,\hg}}{\bf B}_{\sss{\ha\hb}}
  &= i\,\ve\1_{\sss\hK}{}^{\sss\hL}\,
         \big[\,\d\1_{\sss\hg[\ha}\bs{\J}\1_{\sss\hL\,\hb]}
          ~-~ \ve\1_{\sss\ha\hb\hg}{}^{\sss\hd}\,
               \bs{\J}\1_{\sss\hL\,\hd}\,\big] ,
 \label{UM2.2}\\[2mm]
 \rD\1_{\sss\hK\,\ha}\bs{\J}\1_{\sss{\hL\,\hg}}
  &=\hbox to0mm{$\ddd\big[
      \d_{\sss{\ha\hg}}\,(\s^3)_{\sss{\hK\hL}}\,(\ddt{\bf A})
     +\d_{\sss{\hK\hL}}\,(\ddt{\bf A}_{\sss{\ha\hg}})
     +\d_{\sss{\ha\hg}}\,(\s^1)_{\sss{\hK\hL}}\,(\ddt{\bf B})
     +\ve_{\sss{\hK\hL}}\,(\ddt{\bf B}_{\sss{\ha\hg}})\big].$\hss}
 \label{UM2.3}
 \quad&\quad
 &\mkern350mu
\end{alignat}
\end{subequations}
The {\em\/superfield multiplet\/} $({\bf A},{\bf B},{\bf A}_{\sss{\ha\hb}},{\bf B}_{\sss{\ha\hb}}|\bs{\J}_{\sss{\hK\,\ha}})$ constrained by the relations\eq{UM2} describes the ultra-multiplet in terms of {\em\/a priori\/} unconstrained, off-shell superfields, the use of which ought to facilitate eventual quantization by path-integral methods. 
 Given the super-differential equations\eq{UM2}, a direct calculation on all the fields in the multiplet implies that\eq{eSuSyD} are satisfied; each index $I,J,\dots$ therein corresponds to an index-pair $(\hK,\ha),(\hL,\hb),\dots$. As promised in the introduction, this notation exhibits a manifest $\SZ^e$-labeling.

Finally, there is a simple Lagrangian that is invariant with respect to $N=8$ extended worldline supersymmetry, in the usual sense of supersymmetric theories:
\be
 {\cal L}_\text{ultra-mult.}
 ~=~ \fracm12\,(\ddt A)^2
  ~+~ \fracm12\,(\ddt B)^2
  ~+~ \fracm18\,(\ddt A_{\sss{\Hat\a\Hat\g}})^2
  ~+~ \fracm18\,(\ddt B_{\sss{\Hat\a\Hat\g}})^2
  ~-~ i\,\fracm12\,\j^{\sss{\Hat\rK\,\Hat\a}} \ddt\j_{{}_{\Hat\rK\,\Hat\a}}.
 \label{UM4}
\ee
Given here in component form for simplicity, this Lagrangian in fact has a manifestly $N=8$ supersymmetric formulation in superspace\cite{rGLP,r6-2}, as afforded by the superfield multiplet formulation\eqs{UM1.5}{UM2}. Thus,\eq{UM1} and\eq{UM4} describe a local, $N=8$ supersymmetric free-field model for the $(A,B,A_{\sss\ha\hb},B_{\sss\ha\hb}|\j\1_{\sss\hK\ha})$ supermultiplet on the worldline.

More importantly however, this model provides a basis for an entire family of supersymmetric models that are closely related to\eq{UM1} and\eq{UM4}.  These related models may be revealed by using the ``root superfield'' formalism of Ref.\cite{rGLP}.  We can implement this approach here by simply replacing the bosonic component fields---and so also the superfields\eq{UM1.5}---appearing above according to the rules
\be
   A ~\to \ddt^{-a_1} A,\quad
   B ~\to \ddt^{-a_2} B,\quad
   A_{{}_{\Hat\a\Hat\g}} ~\to \ddt^{-a_3} A_{{}_{\Hat\a\Hat\g}},\quad
   B_{{}_{\Hat\a\Hat\g}} ~\to \ddt^{-a_4} B_{{}_{\Hat\a\Hat\g}}, 
 \label{UM5}
\ee
where $a_1$, $a_2$, $a_3$ and $a_4$ are non-negative integers.  From the form of the Lagrangian\eq{UM4}, it is seen to remain local as long as these integers $a_1,\cdots,a_4$ are chosen to be either 0 or 1.  It can also be easily shown that the apparent non-locality introduced into the supersymmetry transformation laws\eq{UM2} is illusionary.  The transformation laws can be rewritten in completely local 
ways after implementation of\eq{UM5}.  We call the process of changing the values of
an exponent from, say, $a_1=0$ to $a_1=1$ ``raising the node $A$.'' This operation has the remarkable property of changing the propagating bosonic field, in this case $A$, into an auxiliary bosonic field:
\be
 {\cal L}^{(1,0,0,0)}_\text{ultra-mult.}
 ~=~ \fracm12\,(\ddt B)^2
  ~+~ \fracm18\,(\ddt A_{{}_{\Hat\a\Hat\g}})^2
  ~+~ \fracm18\,(\ddt B_{{}_{\Hat\a\Hat\g}})^2
  ~-~ i\,\fracm12\,\j^{\sss{\Hat\rK\,\Hat\a}} \ddt\j_{{}_{\Hat\rK\,\Hat\a}}
  ~+~ \fracm12\,A^2.
 \label{UM4.1}
\ee
The entire family of models is enumerated by the 
$\{0,1\}$-valued components of the vector $\vec{a}=(a_1,a_2,a_3,a_4)$.  Another example is provided by the Lagrangian in the case where $\vec{a}=(1,1,1,1)$:
\be
 {\cal L}^{(1,1,1,1)}_\text{ultra-mult.}
 ~=~ -\,i\,\fracm12\,\j^{\sss{\Hat\rK\,\Hat\a}} \ddt\j_{{}_{\Hat\rK\,\Hat\a}}
  ~+~ \fracm12\,A^2
  ~+~ \fracm12\,B^2
  ~+~ \fracm18\,(A_{{}_{\Hat\a\Hat\g}})^2
  ~+~ \fracm18\,(B_{{}_{\Hat\a\Hat\g}})^2,
 \label{UM6}
\ee
where only the fermions are seen to describe any on-shell propagating degrees of freedom.

The symmetries of the formulation of the ultra-multiplet above allow for a remarkable circumstance. In the $(0,0,0,0)$-action of\eq{UM4} there are no auxiliary fields.  In the $(1,1,1,1)$-action of\eq{UM6} there are eight auxiliary fields. The symmetries and structure of the system\eq{UM2} are precisely such that they permit a sequence of models with $1, 2,\cdots,8$ bosonic auxiliary fields to appear.  Table~\ref{t:I} specifies the enumeration of how all these models correspond to the exponents $\vec{a}$. A supersymmetric free Lagrangian of the type\eqs{UM4}{UM4.1}--(\ref{UM6}) may be fashioned easily for each of them by starting from\eq{UM4} and performing the substitution\eq{UM5} according to the desired choice, picked from Table~\ref{t:I}. Each of these has a manifestly supersymmetric rendition, as shown in Ref.\cite{rGLP,r6-2}.
\begin{table}[htbp]
 $$
  \begin{array}{@{} cccc @{}}
  \begin{tabular}{r@{~}l}
   \small\Bf\# &\small\Bf of propagating\\[-2mm]
               &\small\Bf bosonic fields
  \end{tabular}
 &\begin{tabular}{r@{~}l}
   \small\Bf\# &\small\Bf of auxiliary\\[-2mm]
               &\small\Bf bosonic fields
  \end{tabular}
 &\bs{\vec{a}=(a_1,a_2,a_3,a_4)}
 &\begin{tabular}{c}
   \small\Bf Description\\[-2mm]
   \small\Bf Degeneracy
  \end{tabular} \\ 
    \toprule
    8 & 0 & (0,0,0,0) & 1 \\ 
    7 & 1 & (1,0,0,0), (0,1,0,0) & 2 \\ 
    6 & 2 & (1,1,0,0) & 1 \\ 
    5 & 3 & (0,0,1,0),(0,0,0,1) & 2 \\[1mm]
    4 & 4 & ~~\begin{array}{c}
              (1,0,1,0), (0,1,1,0)\\[-1mm]
              (1,0,0,1), (0,1,0,1)
             \end{array}\Big\} & 4 \\[1mm]
    3 & 5 & (1,1,1,0), (1,1,0,1) & 2 \\ 
    2 & 6 & (0,0,1,1) & 1 \\ 
    1 & 7 & (1,0,1,1), (0,1,1,1) & 2 \\ 
    0 & 8 & (1,1,1,1) & 1 \\
    \bottomrule
  \end{array}
 $$
  \caption{Exponents for adapting the Valise ultra-multiplet model\eqs{UM1}{UM4} for node-raising, the numbers of propagating and auxiliary bosons in each, and the degeneracy of this description.}
  \label{t:I}
\end{table}

We will return below to explain the ``description degeneracy'' by way of tracing the group-theoretic reason for the existence of this peculiar $\SZ^e$-basis \cite{rGR0}. First, however, we reconsider the above-described ultra-multiplet family of supermultiplets, using Adinkras\cite{rA,r6-1,r6-3,r6-3.2,r6-3.4}.

\section{Graphic Depictions of the Ultra-Multiplet Family}
 \label{s:E8}
A great many supermultiplets|and certainly the ultra-multiplet as we will see|turn out to be describable in terms of Adinkras\Ft{Owing to the 1--1 correspondence between the superfield multiplet\eqs{UM1.5}{UM2} and the supermultiplet consisting of only the lowest components\eq{UM1.5}, Adinkras such as\eq{eAB88} and\eq{eAB781} are seen to also depict, and just as faithfully, the entire superfield multiplet\eqs{UM1.5}{UM2}. In turn, this implies that every adinkraic supermultiplet has a manifestly supersymmetric off-shell formulation in superspace|{\em\/in addition\/} to that presented in Refs.\cite{r6-1,r6-1.2}; the details of this|and especially a systematic construction of manifestly supersymmetric action functionals to describe the dynamics of superfield multiplets|are beyond our present scope.}:
\begin{enumerate}\itemsep=-1pt\vspace{-3mm}
 \item For every component field draw a node: open for bosons, closed for fermions.
 \item For every pair of component fields, $\f_i$ and $\j_\hj$, obtained from each other by acting with $Q_\rI$, draw an $I$-colored edge connecting the $i^\text{th}$ open node to the $\hj^\text{\,th}$ closed node:
\begin{enumerate}\itemsep=-1pt\vspace{-3mm}
 \item if $Q_\rI(\f_i)=\ddt^\l\j_\hj$ and $Q_\rI(\j_\hj)=i\ddt^{1-\l}\f_i$
       for $\l=0$ or $1$, draw the edge solid;
  \item if $Q_\rI(\f_i)=-\ddt^\l\j_\hj$ and $Q_\rI(\j_\hj)=-i\ddt^{1-\l}\f_i$
       for $\l=0$ or $1$, draw the edge dashed.
\end{enumerate}\vspace{-1mm}
 \item Position the nodes at relative heights that are proportional to the engineering units of the corresponding component fields. Here and in Appendix~\ref{a:D2Q}, we use the symbols $\phi_i$ and $\j_{\hat\jmath}$ as generic labels for the bosons and fermions in ultra-multiplet.
\end{enumerate}\vspace{-3mm}

Using these rules, the ultra-multiplet\eq{UM1} and\eq{UM2} is depicted by the Adinkra:
\begin{equation}
 \vC{\begin{picture}(140,25)
      \put(0,-5){\includegraphics[width=120mm]{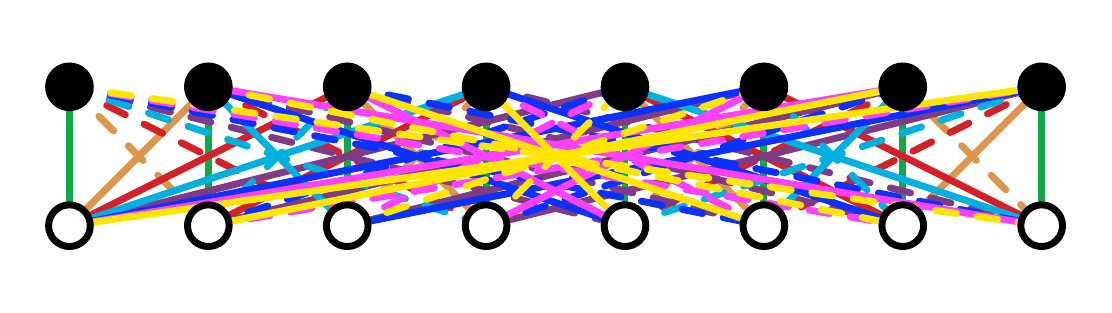}}
     \end{picture}}
 \label{eAB88}
\end{equation}
and the Adinkra of its one-node-raise, for which\eq{UM4} presents a free-field Lagrangian, is:
\begin{equation}
 \vC{\begin{picture}(140,38)
      \put(0,-5){\includegraphics[width=120mm]{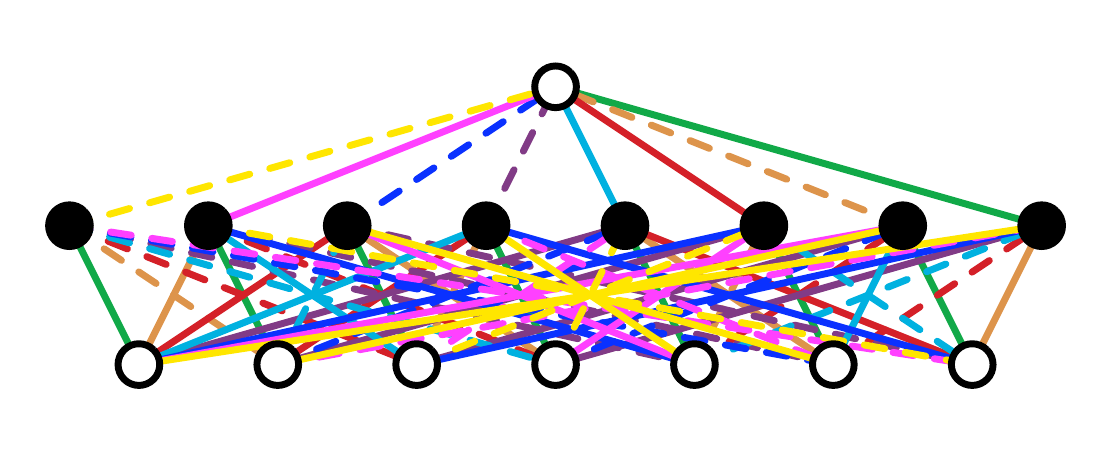}}
     \end{picture}}
 \label{eAB781}
\end{equation}
These depictions of the supermultiplets are faithful: we can reverse-engineer the supersymmetry transformation rules and recover the likes of \Eqs{UM1}{UM2} completely:
 Given any Adinkra, we reverse the above assignments and obtain:
\begin{equation}
 \left. \begin{array}{r@{\>}l}
          Q_{\sss \rI}\,\f_i   &= (\rL_{\sss \rI})_i{}^\hj\,\j_\hj,\\
          Q_{\sss \rI}\,\j_\hj &= i(\rR_{\sss \rI})_\hj{}^k\,(\ddt\f_k);
        \end{array}\right\}\qquad\Iff\qquad
 \left\{\begin{array}{r@{\>}lr@{\>}l}
          \rD_{\sss \rI} \,\bs{\F}_i   &= -i(\rL_{\sss \rI})_i{}^\hj\,\bs{\J}_\hj,\quad
           & \f_i   &:=\bs{\F}_i|,\\
          \rD_{\sss \rI} \,\bs{\J}_\hj &= -(\rR_{\sss \rI})_\hj{}^k\,(\ddt\bs{\F}_k),\quad
           & \j_\hj &:=\bs{\J}_\hj|.
        \end{array}\right.
 \label{eSM=SF}
\end{equation}
where, $\bs{\F}_i$ and $\bs{\J}_\hj$ are {\em\/a priori\/} unconstrained and unprojected superfields. The matrices $\rL_\rI$ and $\rR_\rI$ satisfy\cite{rGR1}
\begin{equation}
 \begin{aligned}
 (\rL_\rI)_i{}^\hj(\rR_\rJ)_\hj{}^k + (\rL_\rJ)_i{}^\hj(\rR_\rI)_\hj{}^k
   &= 2\,\d_i{}^k\\
 (\rR_\rJ)_\hi{}^k(\rL_\rI)_k{}^\hj + (\rR_\rI)_\hi{}^k(\rL_\rJ)_k{}^\hj
   &= 2\,\d_\hi{}^\hj
\end{aligned}~\bigg\}\quad\To\quad
  (\rR_\rI)_\hj{}^i = (\rL_\rI^{-1})_\hj{}^i.
 \label{eLRs}
\end{equation}
It may further be proven that the positive-definite canonical metric $\d_{\rI\rJ}$ occurring in the defining relationships of the worldline supersymmetry\eq{eSuSy} induces positive definite metrics over ${\bf R}_\f$ and ${\bf R}_\j$, which then occurs in the Lagrangians such as\eq{UM4}, \eq{UM4.1} and\eq{UM6}. For these to be invariant with respect to $N=8$ extended supersymmetry,
\begin{equation}
  (\rR_\rI)_\hj{}^i = (\rL_\rI^T)_\hj{}^i,
  \qquad\buildrel{\text{(\ref{eLRs})}}\over\Longrightarrow\qquad
  (\rL_\rI^{-1})_\hj{}^i = (\rL_\rI^T)_\hj{}^i.
 \label{eORL}
\end{equation}
That is, the $\rL_\rI,\rR_\rI$ matrices are orthogonal.

The classification program of Refs.\cite{r6-1,r6-3,r6-3.2,r6-3.4} has been completed for $N\leq28$ and is in progress for higher $N$. The $(8|8)$-dimensional ultra-multiplet turns out to be both minimal, and {\em\/essentially\/} unique|up to permutations of the $Q_\rI$, and given the specification $[\j_\hj]=[\f_i]+\inv2$. The particular ``wiring diagram'' formed by the 8-colored edges of the Adinkra together with the coloring of the nodes is called the {\em\/chromotopology\/} of the Adinkra and is encoded by the $\rL_\rI$ matrices in\eq{eSM=SF}. It may be specified unambiguously as a $(\ZZ_2)^k$-quotient of the 8-cube where the edges along the $I^\text{th}$ dimension are colored in the $I^\text{th}$ color. This quotient nature implies that the following {\em\/quasi\/}-projectors\Ft{These local operators square to an $H^2$-multiple of themselves, rather than themselves. Having noted this, we will for simplicity no longer insert the ``quasi'' prefix. Ref.\cite{r6-1.2} shows how to construct a system of operators corresponding to\eq{PE8}, which impose perhaps more familiar (anti-)self-duality conditions.}:
\begin{subequations}
 \label{PE8}\vspace{-1mm}
\begin{alignat}{5}
 \hat\P_{1234}^\pm&:=\inv2\big[\,H^2 \pm Q_1Q_2Q_3Q_4,\big]
  \qquad&\iff\qquad
 \bs{b}_1&=[\,1~1~1~1~0~0~0~0\,],\\
 \hat\P_{3456}^\pm&:=\inv2\big[\,H^2 \pm Q_3Q_4Q_5Q_6,\big]
  \qquad&\iff\qquad
 \bs{b}_2&=[\,0~0~1~1~1~1~0~0\,],\\
 \hat\P_{5678}^\pm&:=\inv2\big[\,H^2 \pm Q_5Q_6Q_7Q_8,\big]
  \qquad&\iff\qquad
 \bs{b}_3&=[\,0~0~0~0~1~1~1~1\,],\\
 \hat\P_{2468}^\pm&:=\inv2\big[\,H^2 \pm Q_2Q_4Q_6Q_8,\big]
  \qquad&\iff\qquad
 \bs{b}_4&=[\,0~1~0~1~0~1~0~1\,],
\end{alignat}
\end{subequations}
act so as to produce an $H^2$-multiple of each component field, and so do also all their products. The four binary 8-vectors at the right-hand side of\eq{PE8} generate the $e_8$ doubly-even binary linear block code. We say that the the graph\eq{eAB88} is an  $e_8$-encoded quotient, $I^8/(\ZZ_2)^4$, of the 8-cube, $I^8$.
 Visually, this may be verified by tracing a 4-colored quadrangle, in any of the four four-color groups indicated in\eq{PE8} or by any of their products, and starting from any node: such quadrangles always close, and with an overall sign that is determined by the sign-choice in\eq{PE8}.
 
Conceptually, this quotient nature of the minimal supermultiplets of $N=8$ extended supersymmetry is very similar to the familiar decomposition of the 4-compo\-nent Dirac spinor in 4-dimensional spacetime into the left- and right-handed 2-component Weyl spinors. It is just that for $N=8$, there exist {\em\/four\/} mutually commuting projections, implemented, \eg, by the operators\eq{PE8}.

The operators\eq{PE8} do not commute with the infinitesimal supersymmetry transformation operator, $\d_Q(\e):=\e^\rI Q_\rI$, and the {\em\/a priori\/} unconstrained and unprojected $(128|128)$-dimensional supermultiplet
\begin{equation}
 \big(\,\f_{[I_1\cdots I_{2k}]}\>|\>\j_{[I_1\cdots I_{2k+1}]}\,\big),
  \quad
 F_{[I_1\cdots I_k]}
  := H^{-\lfloor\frac{k}2\rfloor}Q_{[I_1}\cdots Q_{I_k]}(F_0)
  =\Big\{\begin{array}{l@{~\text{if}~}l}
           \f_{[I_1,\cdots,I_k]} & k\text{ is even},\\[-1mm]
           \j_{[I_1,\cdots,I_k]} & k\text{ is odd},
          \end{array}
 \label{e128}
\end{equation}
is not left invariant by any of the operators\eq{PE8}. However, it contains sub-supermultiplets that are left invariant by the {\em\/right-action\/} of the operators\eq{PE8}:
\begin{equation}
  \hat\P^\pm_*(F_{[I_1,\cdots,I_k]})
   :=H^{-\lfloor\frac{k}2\rfloor}Q_{[I_1}\cdots Q_{I_k]}\,\hat\P^\pm_*(F_0),
\end{equation}
so that, for example:
\begin{subequations}
 \label{e64+}
 \begin{alignat}{5}
   F^+_0&:=(F_0+F_{1234}),
    &\qquad
   F^+_5&:=F_5+F_{12345},
    &\qquad
   F^+_{25}&:=F_{25}-F_{1345},\\
   F^+_1&:=(F_1+F_{234}),
    &\qquad
   F^+_{12}&:=F_{12}+F_{34},
    &\qquad
   F^+_{125}&:=F_{125}+F_{345},\\
   F^+_2&:=(F_2-F_{134}),
    &\qquad
   F^+_{15}&:=F_{15}+F_{2345},
    &\qquad
   &\text{\etc}
 \end{alignat}
\end{subequations}
The relative signs in the component field definitions in the right-hand side of\eq{e64+} ensure both that each so-defined component field turns into a uniform, $+H^2$-multiple of itself upon the right-action of $\hat\P^+_{1234}$, and also that the resulting, $(64|64)$-dimensional supermultiplet is a proper, closed orbit of the supersymmetry algebra\eq{eSuSy} under the usual, left-action of the $Q$'s:
\begin{subequations}
 \label{eSS64}
 \begin{alignat}{7}
   Q^{}_1F_0^+&=iF_1^+,&\qquad
   Q^{}_2F_0^+&=iF_2^+,&\qquad&\cdots&\qquad
   Q^{}_5F_0^+&=iF_5^+,\\
   Q^{}_1F_1^+&=\ddt F_0^+,&\qquad
   Q^{}_2F_1^+&=-\ddt F_{12}^+,&\qquad&\cdots&\qquad
   Q^{}_5F_1^+&=-\ddt F_{15}^+,\\
   Q^{}_1F_2^+&=\ddt F_{12}^+,&\qquad
   Q^{}_2F_2^+&=\ddt F_0^+,&\qquad&\cdots&\qquad
   Q^{}_5F_2^+&=-\ddt F_{25}^+,\\
   Q^{}_1F_5^+&=\ddt F_{15}^+,&\qquad
   Q^{}_2F_5^+&=\ddt F_{25}^+,&\qquad&\cdots&\qquad
   Q^{}_5F_5^+&=\ddt F_0^+,
 \end{alignat}
\end{subequations}
and so on.
In turn, $\hat\P^-_{1234}$ will {\em\/right-annihilate\/} the so-defined $F_{\cdots}^+$'s; for example,
\begin{subequations}
 \label{e-64+}
 \begin{align}
 \hat\P_{1234}^-(F_0^+)&=\hat\P_{1234}^-(F_0+F_{1234})
  =\hat\P_{1234}^-\big(F_0+H^{-2}Q_1Q_2Q_3Q_4F_0\big),\\
 &=\hat\P_{1234}^-(F_0)+H^{-2}Q_1Q_2Q_3Q_4\,\hat\P_{1234}^-(F_0),\\
 &=\inv2\,\big[H^2-Q_1Q_2Q_3Q_4]F_0
   +\inv2H^{-2}Q_1Q_2Q_3Q_4\big[H^2-Q_1Q_2Q_3Q_4\big]F_0,\nn\\
 &=\inv2\,\big[H^2-(+H^2)\big]F_0,~=~0.
\end{align}
\end{subequations}
That is, $\ker(\hat\P_{1234}^+)=\img(\hat\P_{1234}^-)$. Having thus accomplished the $\hat\P_{1234}^+$-projection, we construct the $\hat\P_{3456}^+$-projection thereof. Upon this $(32|32)$-dimensional $(\ZZ_2)^2$-quotient supermultiplet, the right-action of the product operator
\begin{align}
 \hat\P_{1234}^+\hat\P_{3456}^+
 &=\inv4\big[H^4+H^2Q_1Q_2Q_3Q_4+H^2Q_3Q_4Q_5Q_6-H^2Q_1Q_2Q_5Q_6\big],\nn\\
 &\simeq\inv4\big[H^4-H^4-H^4-H^2Q_1Q_2Q_5Q_6\big]
  =-\inv2H^2\big[H^2+Q_1Q_2Q_5Q_6\big]=-H^2\hat\P_{1256}^+
\end{align}
is indistinguishable from an $H^4$-action.
 Next, we similarly construct the subsequent $\hat\P_{3456}^+$-projection thereof, and finally the $\hat\P_{3456}^+$-projection, resulting in the $(8|8)$-dimensional supermultiplet\eq{eAB88}.

For the system\eq{PE8}, we have for all 4-plets ${\cal I},{\cal K}=1234,\,3456,\,5678,\,2468$:
\begin{equation}
   \hat\P_{\cal I}^+ + \hat\P_{\cal I}^- = H^2,\qquad
   \hat\P_{\cal I}^+ \circ \hat\P_{\cal I}^- = 0,\qquad\text{and}\qquad
   \big[\,\hat\P_{\cal I}^\pm\,,\,\hat\P_{\cal K}^\pm\,\big]=0=
   \big[\,\hat\P_{\cal I}^\pm\,,\,\hat\P_{\cal K}^\mp\,\big],
\label{eP's}
\end{equation}
so that the successive application of any $\hat\P_{\cal I}^{\b_{\cal I}}{\circ}\hat\P_{\cal K}^{\b_{\cal I}}$, for any ${\cal I}\neq{\cal K}$ and $\b_{\cal I},\b_{\cal K}=\pm1$, quarters the supermultiplet\Ft{This situation is not unfamiliar to physicists: the projector to a Majorana-real spinor complements the one to the Majorana-imaginary spinor, and these two annihilate each other. Similarly the Weyl projector to a left-handed spinor complements the one to a right-handed one, and they annihilate each other. In general, there is no reason for a Majorana and a Weyl projector to satisfy any relation; but if they commute, we can construct Majorana-Weyl spinors that have a quarter of the degrees of freedom of a Dirac spinor.}. The application of any
\begin{equation}
 \hat\P_{\cal I}^{\b_{\cal I}}\,\hat\P_{\cal J}^{\b_{\cal J}}\,
  \hat\P_{\cal K}^{\b_{\cal K}}\,\hat\P_{\cal L}^{\b_{\cal L}},\qquad
  \text{with the multi-indices $\cal I,J,K,L$ all different}
 \label{e16E8Q}
\end{equation}
then cuts the component field content of the {\em\/a priori\/} unconstrained and unprojected supermultiplet to its $16^\text{th}$: $(128|128)\to(8|8)$. $N=8$ is the lowest number of worldline supersymmetries, $N$, for which this maximal $2^{N/2}$-fold reduction, through a $(\ZZ_2)^{N/2}$-quotient, can occur.

The possible choices of the four relative signs, $\b_{\cal I}$, in the operators\eq{PE8} provide $2^4=16$ distinct projections. However, quotient supermultiplets with the same product $\prod_{\cal I}\b_{\cal I}$ are equivalent to each other by simple field redefinitions, as detailed in Construction~4.2 of Ref.\cite{r6-3.2}. In turn, no field redefinition can transform a member of the  $\prod_{\cal I}\b_{\cal I=}+1$ equivalence class into any member of the  $\prod_{\cal I}\b_{\cal I}=-1$ equivalence class. Generalizing the nomenclature of Ref.\cite{rGHR}, we use
\begin{defn}\label{dUM}
With the notation as in Eqs.\eq{PE8}, \eq{e128} and\eq{e16E8Q}, a Valise $(8|8)$-dimensional $(\ZZ_2)^4$-quotient supermultiplet of $N=8$-extended worldline supersymmetry with $\prod_{\cal I}\b_{\cal I}=+1$ is an {\bfseries ultra-multiplet}, one with $\prod_{\cal I}\b_{\cal I}=-1$ is a {\bfseries twisted ultra-multiplet}. Their node-raised relatives then populate, respectively, the ultra-multiplet family and its twisted variant.
\end{defn}

All such iterated projections turn out to be classified by doubly-even linear binary block codes, used for error-detecting and error-correcting in information transfer. The permutation equivalence class of codes corresponding to the quartet of projection operators\eq{PE8} is denoted $e_8$, and is indeed related to the familiar $E_8$ Lie group and corresponding lattice\cite{r6-3}; see Appendix~\ref{a:E8}. The computation that unambiguously determines whether two Adinkras and their corresponding supermultiplets are equivalent involves the $\ZZ_2$-valued cubical cohomology of the Adinkras, as detailed in Ref.\cite{r6-3.4}.

\section{Symmetries in the Ultra-Multiplet Family}
 \label{s:SymUM}
Comparing the Adinkras\eq{eAB88} and\eq{eAB781}, it is evident that the right-most open node was raised from the bottom row of the former Adinkra to obtain the latter. This node should thus be identified with either one of the ``singlet'' component fields $A,B$, when passing from the Lagrangian\eq{UM4.1} to\eq{UM6}. Considering the Adinkra\eq{eAB88} however, it seems self-evident that the eight open nodes offer completely equivalent candidates for raising: Any one of them could have been raised and so identified with either one of $A,B$, upon which a corresponding permutation of the remaining nodes and a redefinition of some of the component fields into their own negatives (which swaps the dashedness of every incident edge) would render the result indistinguishable from\eq{eAB781} and\eq{UM4.1}.

Indeed, there is a major difference between the specification\eqs{UM1}{UM2} and the supermultiplets described in Section~\ref{s:E8}:
\begin{enumerate}\vspace{-3mm}\itemsep=-1pt
 \item The specifications\eqs{UM1}{UM2} manifestly admit a continuous group of symmetries, $\SZ$, and so describe an inherently continuous equivalence class of supermultiplets.
 \item The projections\eq{PE8} break the $\textsl{O}(8)$ symmetry of\eq{eSuSy} to a subgroup $\D(e_8)$, and so describe an equivalence class of objects corresponding to the (discrete) graphs called Adinkras.
\end{enumerate}\vspace{-3mm}
It may be shown that $|\D(e_8)|=\frac{8!}{1344}=30$\cite{r6-3}, so that there is a total of 30 distinct, but $Q$-permutation equivalent systems of projectors such as\eq{PE8}; see also Appendix~\ref{a:Same}. Each one of these 30 classes of permutations of the system\eq{PE8} defines a 1344-component equivalence class of supermultiplets depicted as\eq{eAB88}. In each of these, a basis of $Q_1,\cdots,Q_8$ is fixed, which in turn ties the fermions' basis rigidly to the basis of the bosons.

\paragraph{Valise Symmetry:}
We may therefore turn this around and ask for the {\em\/most general\/} linear redefinitions of the bosonic component fields, $\f_i\to\Tw\f_i$, the fermionic fields, $\j_\hj\to\Tw\j_\hj$, and the supercharges|or, analogously, the superfields and the super-differential operators in\eqs{UM1.5}{UM2}, the result of which would still furnish an $(8|8)$-dimensional supermultiplet of the $N=8$ extended worldline supersymmetry\eq{eSuSy} and with $[\Tw\j_\hj]=[\Tw\f_i]+\inv2$. The adinkraic representatives in this continuous family of supermultiplets will, by the classification of Refs.\cite{r6-3,r6-3.2}, have to be $Q$-permutation equivalent supermultiplets, with $E_8$ topology, and depicted as\eq{eAB88}|these being unique adinkraic $(8|8)$-dimensional supermultiplets of $N=8$ worldline supersymmetry. For similar reasons, the supermultiplets specified in \Eqs{UM1}{UM2} would also have to find their home in this {\em\/maximal\/} continuous family.

Members in this family may be partitioned through a hierarchy of increasingly subtler distinctions, and Appendix~\ref{a:Same} details some of the possible types of isomorphisms and ensuing equivalence classes. Suffice it here to state that we clearly distinguish between the ultra-multiplet and its twisted variant, as stated in definition~\ref{dUM}.

We will refer to such most general basis-redefining transformations as {\em\/effective symmetries\/}, and denote their group by $\Gf$. We reserve the term {\em\/dynamical symmetries\/} for the analogous notion specified by the dynamics, \ie, action functionals, and note that they are logically separate from $\Gf$, which is determined entirely from the structure of the supermultiplet itself.

Since the eight bosons $\f_i$ have identical engineering units, we may as well consider their arbitrary real linear combinations; the same is true of the fermions. We therefore require that
\begin{equation}
  {\bf R}_Q:=\Span(Q_1,\cdots,Q_8),\quad
  {\bf R}_\f:=\Span(\f_1,\cdots,\f_8),\quad
  {\bf R}_\j:=\Span(\j_1,\cdots,\j_8)
 \label{eRQfj}
\end{equation}
are all real 8-dimensional representations of $\Gf$. In addition, as mentioned in the discussion of \Eqs{eLRs}{eORL}, there exist canonical positive-definite metrics on ${\bf R}_Q$, ${\bf R}_\f$ and ${\bf R}_\j$. Preserving these, it must be that $\Gf\subset\textsl{O}(8)_Q\times\textsl{O}(8)_\f\times\textsl{O}(8)_\j$.

Also, the $\Gf$-representation assignments\eq{eRQfj}|and eventually\eq{e888}|must agree with the supersymmetry transformation rules\eq{eSM=SF}, so that $\Gf$ is the maximal group with respect to which the assignment\eq{eRQfj} is consistent with the projections
\begin{subequations}
 \label{eReff}
 \begin{equation}
  \hat\p_\j\big({\bf R}_Q\otimes{\bf R}_\f\big)={\bf R}_\j
    \quad\text{and}\quad
   \hat\p_\f\big({\bf R}_Q\otimes{\bf R}_\j\big)={\bf R}_\f
\end{equation}
being of maximum rank, and satisfying the supersymmetry relations\eq{eSuSy}:
\begin{alignat}{3}
 \hat\p_\j\big({\bf R}_Q\otimes\hat\p_\f({\bf R}_Q\otimes{\bf R}_\j)\big)
 &=\hat\p_{\bf1}(\Sym^2{\bf R}_Q)\otimes{\bf R}_\j&&=\Ione\otimes{\bf R}_\j,\\
 \hat\p_\f\big({\bf R}_Q\otimes\hat\p_\j({\bf R}_Q\otimes{\bf R}_\f)\big)
 &=\hat\p_{\bf1}(\Sym^2{\bf R}_Q)\otimes{\bf R}_\f&&=\Ione\otimes{\bf R}_\f,
\end{alignat}

\end{subequations}where $\Sym^2{\bf R}_Q\supset\Ione$ is consistent with ${\bf R}_H=\Ione$ on the worldline\eq{eSuSy}\Ft{This condition changes significantly in more than 1-dimensional {\em\/spacetime\/}, where the right-hand side of this hallmark relation of supersymmetry furnishes a nontrivial representation of the Lorentz symmetry.}.

 To satisfy these requirements, we assign in the notation of Ref.\cite{rSsky}:
\begin{equation}
  {\bf R}_Q:={\bf8}_v,\quad
  {\bf R}_\f:={\bf8}_s,\quad
  {\bf R}_\j:={\bf8}_c\qquad\text{of }\Spin(8).
 \label{e888}
\end{equation}
Returning to the Valise Adinkra\eq{eAB88}, we read off the $(\rL_I)_i{}^\hj$ and $(\rR_\rI)_\hj{}^i$ matrices (see Appendix~\ref{a:D2Q}), and note that they are analogous to the familiar Pauli matrices $(\s^m)_{\a\dot\b}$ and $(\bar\s^m)_{\a\dot\b}$|the off-diagonal blocks in the chiral representation of the Dirac gamma matrices in $3{+}1$-dimensional spacetime.
 The Pauli matrices are invariant with respect to the {\em\/simultaneous\/} Lorentz group $\Spin(1,3)$-transformation of the vector, spinor and co-spinor representations, the elements of which are labeled by the indices $m,\a,\dot\b$, respectively.
 
The analogous computation here proves that the matrices $(\rL_\rI)_i{}^\hj$ and $(\rR_\rI)_\hj{}^i$ are invariant with respect to a simultaneous $\Spin(8)$-transformation of the representations\eq{e888}, the elements of which are labeled by the indices $\rI,i,\hj$, respectively.

So that ${\bf R}_\f$ and ${\bf R}_\j$ in\eq{e888} would be faithful representations of $\Gf$, we must in fact use $\Gf=\Spin(8)$ rather than $\SO(8)$. In fact, we may extend $\Gf=\Spin(8)$ to $\Pin(8)$, by including linear transformations of determinant $-1$, generated by reflections $Q_\rI\to-Q_\rI$, for any odd subset of $\rI=1,\cdots,8$. We recall that the ${\bf R}_s$ and ${\bf R}_c$ in all orthogonal groups are spanned by root-lattice vectors of the form $(\pm\inv2,\cdots,\pm\inv2)$: ${\bf R}_s$ with a positive product of components, and ${\bf R}_c$ with a negative one. It follows that the $\Pin(8)/\Spin(8)\simeq\ZZ_2$ reflections swap the two spinors of $\Spin(8)$: ${\bf8}_s\iff{\bf8}_c$, and are the $\ZZ_2$ part of the $S_3$ outer automorphism of $\Spin(8)$.

\paragraph{Non-Valise Symmetry:}
As the bosonic nodes are raised one by one, $\Gf$ is broken to its subgroups, while maintaining the relations\eq{eReff}, and with ${\bf R}_\f$ decomposing as
\begin{equation}
  \dim({\bf R}_\f) =\bf 8 ~\to~ 7+1 ~\to~ 6+2 ~\to~ 5+3 ~\to~ 4+4 ~\to~ 3+5 ~\to~\cdots
 \label{eSeq}
\end{equation}
Of course, ${\bf R}_\j$ and ${\bf R}_Q$ may well decompose along the way, but this is not evident from the Adinkra. The results are shown in Table~\ref{t:E8}.
\begin{table}[htbp]\small
  \centering
  \begin{tabular}{@{} c@{~~}c@{~}c@{~}c@{~}c@{~}c @{}}
    \toprule
 {\Bf Adinkra} & \boldmath$G_\textbf{eff}$ & \boldmath$Q_{\sss\rI}$
               & \boldmath$\f_i$ & \boldmath$\j_\hi$ & \boldmath$G_\textbf{out}$ \\ 
    \midrule
 \multirow{3}[5]{50mm}{\includegraphics[width=50mm]{Pix/E8B88.pdf}}
 \\ 
& $\Spin(8)$ & ${\bf8}_v$
 & ${\bf8}_s$
 & ${\bf8}_c$ & $S_3$ \\ \\
 \multirow{4}{50mm}{\includegraphics[width=50mm]{Pix/E8B781.pdf}}
 \\ \\
& $\Spin(7)$
 & ${\bf8}$
 & ${\bf7}\oplus{\bf1}$
 & ${\bf8}$ & | \\ \\
 \multirow{4}{50mm}{\includegraphics[width=50mm]{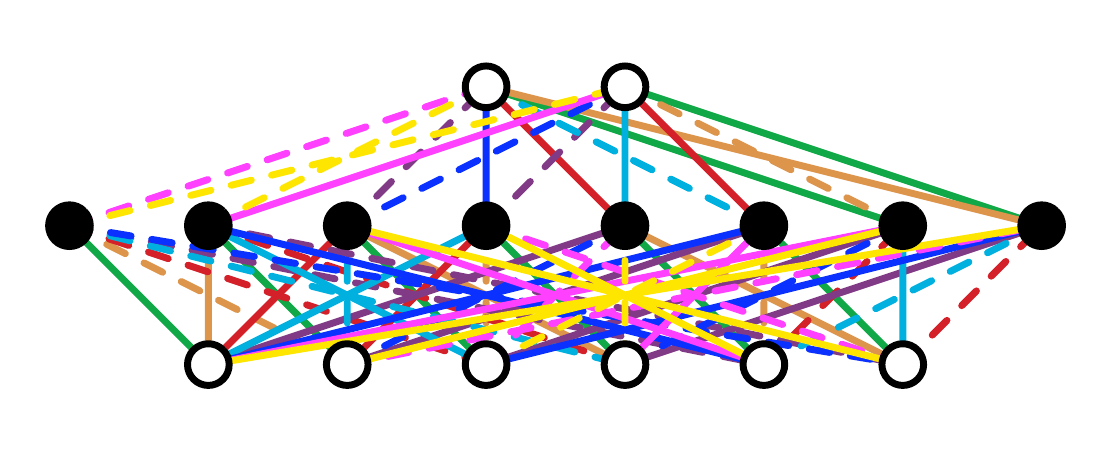}}
 \\ \\
 & $\Spin(6)\times\Spin(2)$
 & ${\bf4}_{+1}\oplus\bs{4^*}\!_{-1}$
 & ${\bf6}_0\oplus{\bf1}_{-2}\oplus{\bf1}_{+2}$
 & ${\bf4}_{-1}\oplus\bs{4^*}\!_{+1}$ & $\ZZ_2^{~2}$ \\
 &  ($\SU(4)\times\textsl{U}(1)$) \\
 \multirow{4}{50mm}{\includegraphics[width=50mm]{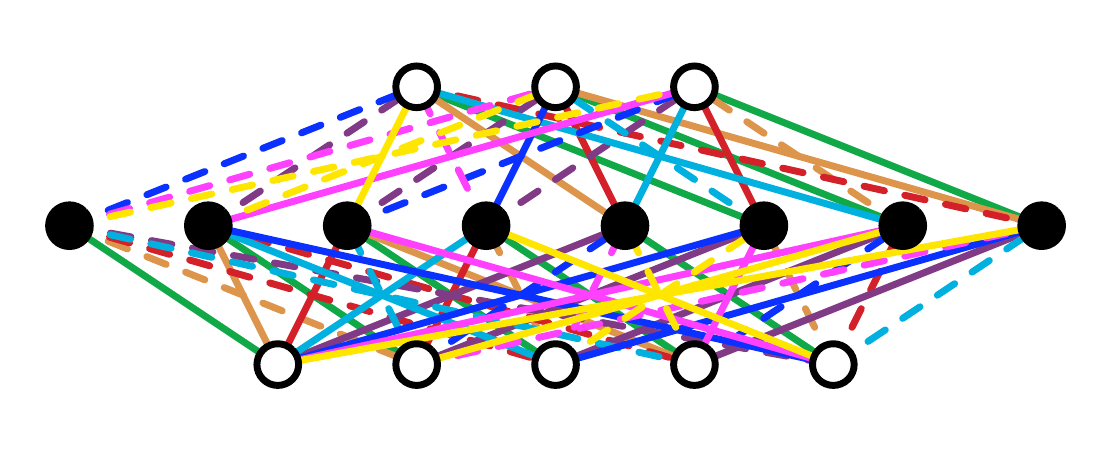}}
 \\ \\
& $\Spin(5)\times\Spin(3)$
 & $({\bf4},{\bf2})$
 & $({\bf5},{\bf1})\oplus({\bf1},{\bf3})$ 
 & $({\bf4},{\bf2})$ & | \\
 & ($\Sp(4)\times\SU(2)$) \\
 \multirow{4}{50mm}{\includegraphics[width=50mm]{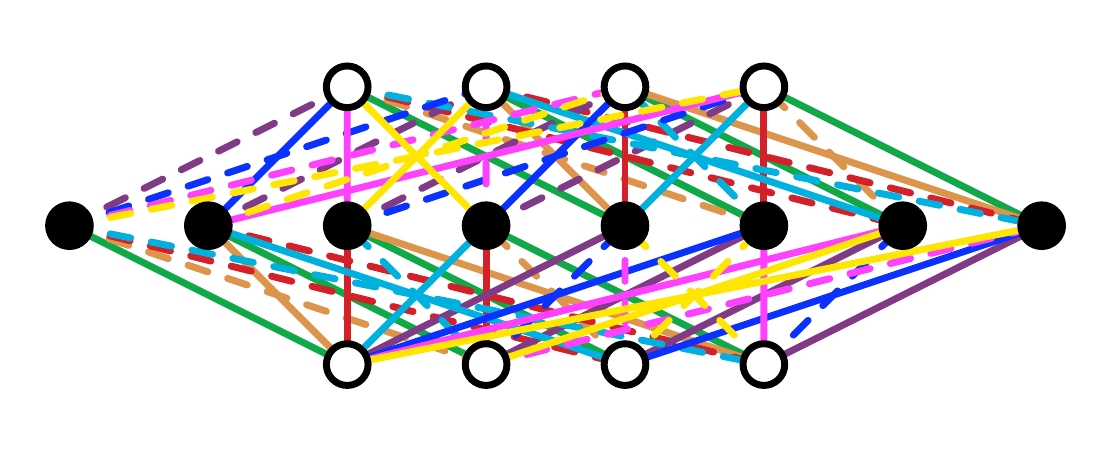}}
 \\ \\
& $\Spin(4)\times\Spin(4)$
 & $\quad({\bf2},{\bf1};{\bf2},{\bf1})$
 & $\quad({\bf2},{\bf2};{\bf1},{\bf1})$ 
 & $\quad({\bf1},{\bf2};{\bf2},{\bf1})$ & $\ZZ_2^{~3}$ \\
 &\hbox to0pt{\hss($\SU(2)^2\times\SU(2)^2$)\hss}
 & $\quad\oplus({\bf1},{\bf2};{\bf1},{\bf2})$
 & $\quad\oplus({\bf1},{\bf1};{\bf2},{\bf2})$ 
 & $\quad\oplus({\bf2},{\bf1};{\bf1},{\bf2})$ & $\ZZ_2^{~3}$ \\
    \bottomrule
  \end{tabular}
  \caption{\small A portion of the family of Adinkras with the $E_8$ chromotopology, the maximal effective symmetry group $G_\text{max}$. Note:
  $\Spin(6)=\SU(4)$,
  $\Spin(5)=\Sp(4)$,
  $\Spin(4)=\SU(2)^2$,
  $\Spin(3)=\SU(2)$,
  $\Spin(2)=\textsl{U}(1)$ and
  $\Spin(1)=\ZZ_2$.
By raising more than four open nodes, the resulting Adinkras look like the ones depicted, but drawn upside-down, and the entires in the remaining columns turn out the same as already shown.}
  \label{t:E8}
\end{table}

\begin{table}[htbp]\footnotesize
 $$
  \begin{array}{@{} llll @{}}
 \bs{\textsl{\bfseries Spin}(8)} & \bs{R_\f=8_s} & \bs{R_Q=8_v} & \bs{R_\j=8_c} \\ 
  \toprule
 \Spin(7)
  & {\bf7}\oplus{\bf1}
  & {\bf8} & {\bf8} \\
 \to\Spin(6)=\SU(4)
  & ({\bf6}\oplus{\bf1})\oplus{\bf1}
  & {\bf4}\oplus\bs{4^*}
  & {\bf4}\oplus\bs{4^*} \\
 ~\too{\ddag1}\SZ^e
  & (\6({\bf1,3})_0\oplus\6({\bf3,1}))_0\oplus({\bf1,1})_0\oplus({\bf1,1})_0
  & ({\bf2,2})_+\oplus({\bf2,2})_-
  & ({\bf2,2})_-\oplus({\bf2,2})_+ \\
   \midrule
 \Spin(6){\times}\Spin(2)
  & {\bf6}_0\oplus{\bf1}_{-2}\oplus{\bf1}_{+2}
  & {\bf4}_{+1}\oplus\bs{4^*}\!_{-1}
  & {\bf4}_{-1}\oplus\bs{4^*}\!_{+1} \\
 ~\too{\ddag2}\SZ^e
  & (\6({\bf1,3})_0\oplus\6({\bf3,1}))_0\oplus({\bf1,1})_0\oplus({\bf1,1})_0
  & ({\bf2,2})_+\oplus({\bf2,2})_-
  & ({\bf2,2})_-\oplus({\bf2,2})_+ \\
   \midrule
 \Spin(5){\times}\Spin(3)
  & ({\bf5,1})\oplus({\bf1,3})
  & ({\bf4,2})
  & ({\bf4,2}) \\
 \too{\ddag3}\Spin(3)^2{\times}\Spin(2)
  & ({\bf3,1})_0\oplus({\bf1},{\bf1})_{-2}\oplus({\bf1},{\bf1})_{+2}\oplus({\bf1},{\bf3})_0
  & ({\bf2,2})_{+1}\oplus({\bf2,2})_{-1}
  & ({\bf2,2})_{+1}\oplus({\bf2,2})_{-1} \\
 ~\too{\ddag2}\SZ^e
  & (\6({\bf1,3})_0\oplus\6({\bf3,1}))_0\oplus({\bf1,1})_0\oplus({\bf1,1})_0
  & ({\bf2,2})_+\oplus({\bf2,2})_-
  & ({\bf2,2})_-\oplus({\bf2,2})_+ \\
   \midrule
 \Spin(4){\times}\Spin(4)
  & ({\bf2,2;1,1})\oplus({\bf1,1;2,2})
  & ({\bf2,1;2,1})\oplus({\bf1,2;1,2})
  & ({\bf1,2;2,1})\oplus({\bf2,1;1,2}) \\
 ~\too{\ddag2}(\SU(2)_{\sss D})^2{\times}\ZZ_2^e
  & \big(({\bf3;1})_0\oplus({\bf1;1})_0\big)\oplus
     \big(({\bf1;3})_0\oplus({\bf1;1})_0\big)
  & ({\bf2;2})_+\oplus({\bf2;2})_-
  & ({\bf2;2})_-\oplus({\bf2;2})_+ \\
    \bottomrule
\multicolumn{4}{l}{\text{Note:}~
                        \Spin(6){=}\SU(4),~
                        \Spin(5){=}\Sp(4),~
                        \Spin(4){=}\Spin(3)^2,~
                        \Spin(3){=}\SU(2),~
                        \Spin(2){=}\textsl{U}(1)~\text{and}~
                        \Spin(1){=}\ZZ_2.}\\
\multicolumn{4}{l}{{}^{\ddag1}\,\ZZ_2^e\text{ labels conjugate spinors ``$\pm1$'' and tensors ``0,'' but only $\ZZ_2$ with ``$+1$''$\cong$``$-1$'' is the symmetry group.}}\\
\multicolumn{4}{l}{{}^{\ddag2}\,\text{$\Spin(2)\to\ZZ_2^e$; see Appendix~\ref{a:Spin}.}\quad {}^{\ddag3}\,\Spin(5){\times}\Spin(3)\to
 \big(\Spin(3){\times}\Spin(2)\big){\times}\Spin(3)=\Spin(3)^2{\times}\Spin(2).}\\
  \end{array}
 $$
  \caption{The Lie group $\Spin(8)$ and its subgroups for which one of ${\bf8}_v$, ${\bf8}_s$ or ${\bf8}_c$ decomposes according to the sequence\eq{eSeq}; adapted from Ref.\protect\cite{rSsky} using outer automorphisms.}
  \label{t:O8}
\end{table}

In all the subgroups listed in Table~\ref{t:O8}, the {\em\/tensorial\/} ${\bf8}_v$ column is assigned to $\Span(Q_1,\cdots,Q_8)$. However, using the $\Spin(8)$ triality, this decomposes into the {\em\/spinorial\/} representations of the subgroups of $\Spin(8)$. In turn, this assignment induces a $\ZZ_2$ action inherited from $\Spin(8)$, with respect to which ${\bf R}_\f=\Span(\f_1,\cdots,\f_8)$ and ${\bf R}_\j=\Span(\j_1,\cdots,\j_8)$ transform as odd (spinorial) representations and $\Span(Q_1,\cdots,Q_8)$ is even (tensorial). In addition, this tracing permits us to distinguish spinors from their conjugates, leaving us with the above-defined $\ZZ_2^e$ throughout Table~\ref{t:O8}. Nevertheless, only $\Spin(1)=\ZZ_2$, which ignores the distinction between the two conjugate spinors of $\Spin(2)$, is a subgroup in each effective symmetry group, $\Gf$.

Finally, once these group-theoretic assignments have been made, we can re-draw the Adinkras more simply, using this $\Gf$-encoded information:
\begin{alignat}{3}
 \vC{\includegraphics[width=100mm]{Pix/E8B88.pdf}}
 &\quad\mapsto\qquad
 \vC{\begin{picture}(30,30)
      \put(-3,4){\includegraphics[width=10mm]{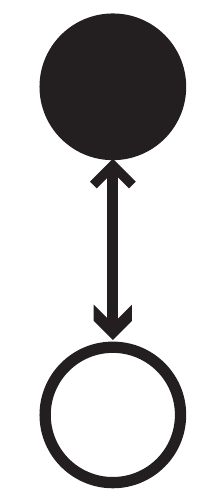}}
      \put(0,7){${\bf8}_s$}
      \put(3,14){${\bf8}_v$}
      \put(0,21){{\color{white}${\bf8}_c$}}
     \end{picture}}
 \label{eO8}\\
 \vC{\includegraphics[width=100mm]{Pix/E8B781.pdf}}
 &\quad\mapsto\qquad
 \vC{\begin{picture}(30,40)
      \put(-2.75,4.5){\includegraphics[width=8mm]{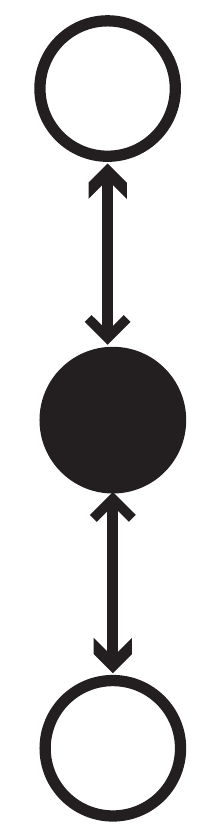}}
      \put(0,6){${\bf7}$}
      \put(2.5,12){${\bf8}$}
      \put(0,18){{\color{white}${\bf8}$}}
      \put(2.5,24){${\bf8}$}
      \put(0,30){${\bf1}$}
     \end{picture}}
 \label{eO7}\\
 \vC{\includegraphics[width=100mm]{Pix/E8B682.pdf}}
 &\quad\mapsto\qquad
 \vC{\begin{picture}(30,40)
      \put(-2.75,4.5){\includegraphics[width=26mm]{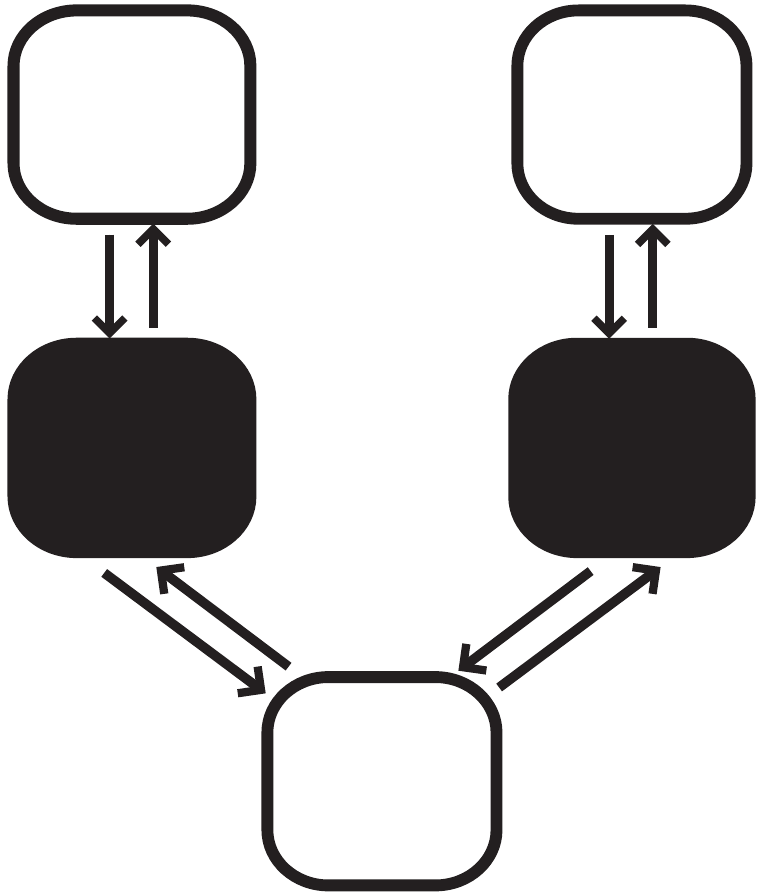}}
      \put(8,7){${\bf6}_0$}
      \put(-1,10){\small${\bf4}_{+1}$}
      \put(7,14){\small$\bs{4^*}\!\!_{-1}$}
      \put(16,10){\small${\bf4}_{+1}$}
      \put(-1.25,19){{\color{white}${\bf4}_{-1}$}}
      \put(15.25,19){{\color{white}$\bs{4^*}\!_{+1}$}}
      \put(-6,24.5){\small${\bf4}_{+1}$}
      \put(7,24.5){\small$\bs{4^*}\!\!_{-1}$}
      \put(21,24.5){\small${\bf4}_{+1}$}
      \put(-1,30){${\bf1}_{-2}$}
      \put(15.5,30){${\bf1}_{+2}$}
     \end{picture}}
 \label{eO6}
\intertext{where the ``inner'' arrows denote successive application of $\bs{4^*}\!\!_{-1}$, whereas the outer arrows denote successive  application of ${\bf4}_{+1}$.}
 \vC{\includegraphics[width=100mm]{Pix/E8B583.pdf}}
 &\quad\mapsto\qquad
 \vC{\begin{picture}(30,40)
      \put(-2.75,4.5){\includegraphics[width=13mm]{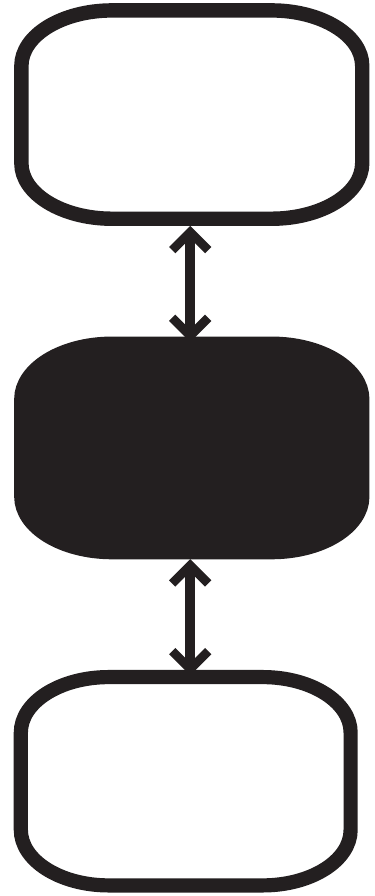}}
      \put(-1.25,7.5){$({\bf5,1})$}
      \put(7,13){$\SSS({\bf4,2})$}
      \put(-1.25,19){{\color{white}$({\bf4,2})$}}
      \put(7,25){$\SSS({\bf4,2})$}
      \put(-1.25,30.5){$({\bf1,3})$}
     \end{picture}}
 \label{eO5}\\
 \vC{\includegraphics[width=100mm]{Pix/E8B484.pdf}}
 &\quad\mapsto\qquad
 \vC{\begin{picture}(30,40)
      \put(-5,4.5){\includegraphics[width=43mm]{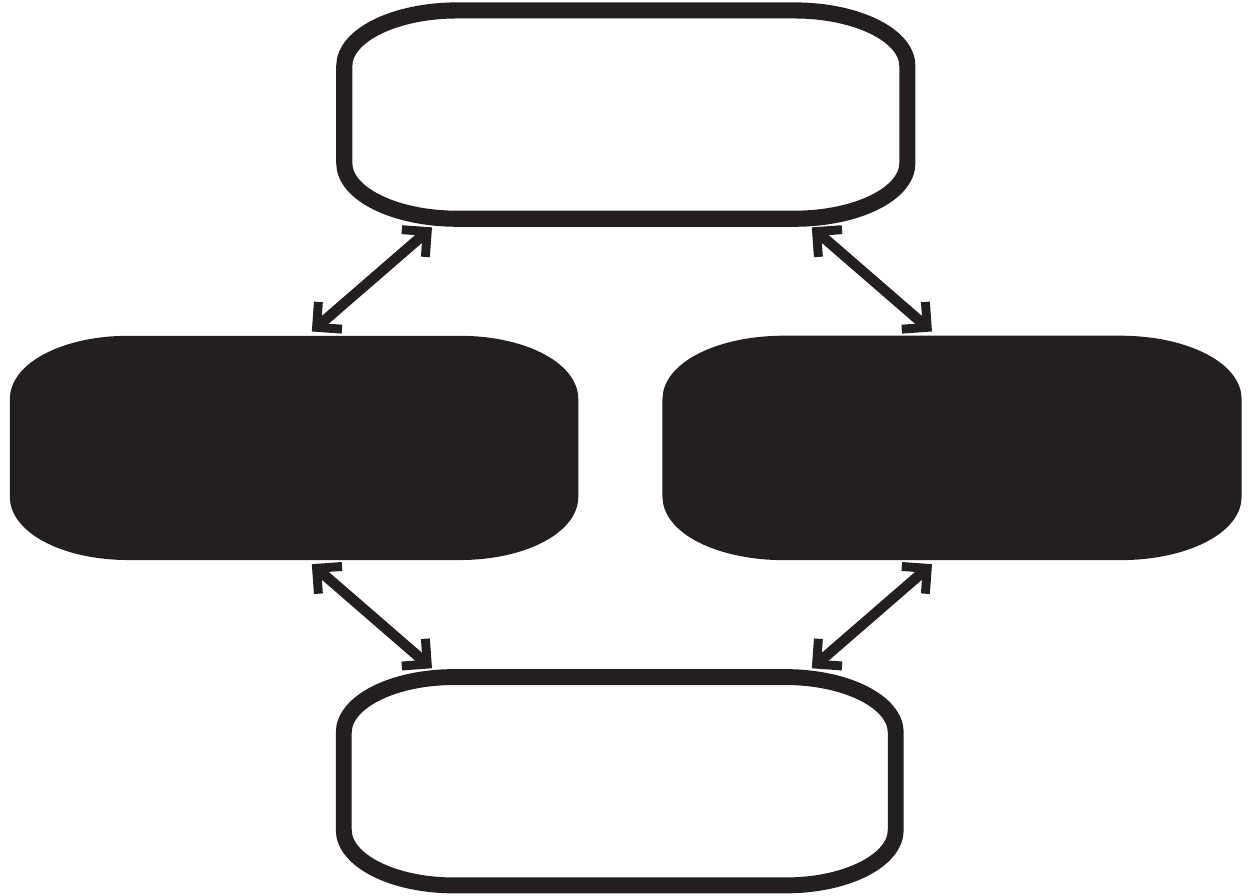}}
      \put(7.25,7.5){$({\bf2,2;1,1})$}
      \put(-5,12){$\SSS({\bf2,1;2,1})$}
      \put(26,12){$\SSS({\bf1,2;1,2})$}
      \put(-4,19){\small{\color{white}$({\bf1,2;2,1})$}}
      \put(19,19){\small{\color{white}$({\bf2,1;1,2})$}}
      \put(26,26){$\SSS({\bf2,1;2,1})$}
      \put(-5,26){$\SSS({\bf1,2;1,2})$}
      \put(7.25,30.5){$({\bf1,1;2,2})$}
     \end{picture}}
 \label{eO4}
\end{alignat}
The Adinkras with $r>4$ raised nodes are upside-down renditions of the ones with $8{-}r$ raised nodes; we thus omit them. By the ultra-multiplet family we herein mean the collection of a total of nine supermultiplets, from the original, $(8|8)$-dimensional ultra-multiplet\eq{eO8}, through\eqs{eO7}{eO4} and on, until all eight bosonic nodes have been raised.

The so-obtained nine distinct supermultiplets are exactly indicated by the first column in Table~\ref{t:I}, and correspond precisely to the concept of a ``root superfield,'' with its specification of the $\vec a$-vector introduced in Ref.\cite{rGLP}.

Once the supermultiplet
\begin{equation}
 \vC{\includegraphics[width=100mm]{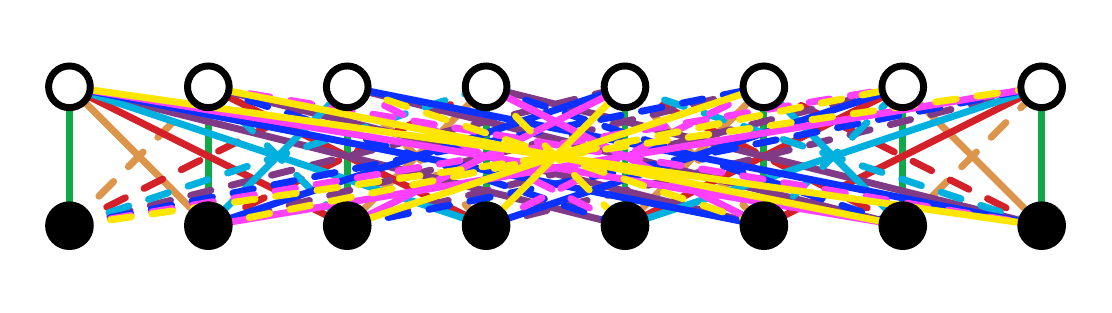}}
 \quad\mapsto\qquad
 \vC{\begin{picture}(30,30)
      \put(-3,26){\rotatebox{180}{\includegraphics[width=10mm]{Pix/O8.pdf}}}
      \put(0,21){${\bf8}_s$}
      \put(3,14){${\bf8}_v$}
      \put(0,7){{\color{white}${\bf8}_c$}}
     \end{picture}}
 \label{eAF88}
\end{equation}
in this family has been reached, we are free to raise one or more of the fermionic (closed) nodes. However, in order to keep the Lagrangians local, this operation forces the introduction of auxiliary fermions into the Lagrangians\eqs{UM4}{UM4.1}--(\ref{UM6}), as discussed in Ref.\cite{rGLP}. We defer revisiting this ``dark side'' of the ultra-multiplet family to a future opportunity, but note that this requires that now ${\bf R}_\j$ decomposes following the pattern\eq{eSeq}.

This suggests swapping the r\^oles of the bosons and the fermions the equations\eqs{UM1}{UM2}, also called a Klein-flip. This may be seen as synonymous with swapping $({\bf R}_\f|{\bf R}_\j)=({\bf R}_s|{\bf R}_c)\to({\bf R}_c|{\bf R}_s)$, which was in turn shown above to be a $\Pin(8)/\Spin(8)\simeq\ZZ_2$ operation generated by reflections $Q_\rI\to-Q_\rI$, and which swap the ultra-multiplet with its twisted variant; see definition~\ref{dUM}. Thus, in the representation\eqs{UM1}{UM2}, the twisting from the definition~\ref{dUM} is equivalent to a Klein-flip followed by the lowering of eight bosonic nodes:
\begin{equation}
 \begin{array}{c@{~}c@{~}cc@{~}l}
 ({\bf A},{\bf B},{\bf A}_{\sss\ha\hb},{\bf B}_{\sss\ha\hb}|\bs\J_{\sss\hK\,\ha})
  &\buildrel{\text{K.\,fl.}}\over\longleftrightarrow&
 (\Tw{\bs\J}{}^+,\Tw{\bs\J}{}^+_{\sss\ha\hb},
   \Tw{\bs\J}{}^-,\Tw{\bs\J}{}^-_{\sss\ha\hb}
  |{\bf C}_{\sss\hK\,\ha}),\\
 &&\doo{\text{8-node lowering}}\\
 &&(\Tw{\bf A}_{\sss\hK\,\ha}|
     \Tw{\bs\J}{}^+,\Tw{\bs\J}{}^+_{\sss\ha\hb},
      \Tw{\bs\J}{}^-,\Tw{\bs\J}{}^-_{\sss\ha\hb}),
  &\text{where}&{\bf C}_{\sss\hK\,\ha}=(\ddt{\Tw{\bf A}}_{\sss\hK\,\ha}).
 \end{array}
 \label{eUM2TUM}
\end{equation}
From the above discussion,
 $(\Tw{\bs\J}{}^+,\Tw{\bs\J}{}^+_{\sss\ha\hb},
    \Tw{\bs\J}{}^-,\Tw{\bs\J}{}^-_{\sss\ha\hb}
  |{\bf C}_{\sss\hK\,\ha})$
is suitable for the ``dark side'' of the ultra-multiplet family, whereas
 $(\Tw{\bf A}_{\sss\hK\,\ha}|
     \Tw{\bs\J}{}^+,\Tw{\bs\J}{}^+_{\sss\ha\hb},
      \Tw{\bs\J}{}^-,\Tw{\bs\J}{}^-_{\sss\ha\hb})$
it the twisted variant of
 $({\bf A},{\bf B},{\bf A}_{\sss\ha\hb},{\bf B}_{\sss\ha\hb}|\bs\J_{\sss\hK\,\ha})$.

We reiterate that each of the so-obtained nine supermultiplets\eq{eO8} and\eq{eAF88} has a free-field local Lagrangian, modeled on\eqs{UM4}{UM4.1}--(\ref{UM6}) and exhibiting $N=8$ extended supersymmetry.

\section{Group-Theoretic Underpinnings}
 \label{s:Roots}
We now turn to the general group-theoretic rationale behind the existence of the $\SZ^e$-basis, resolve the apparent counting discrepancy between the descriptions in Sections~\ref{s:UM} and~\ref{s:E8}, and also find a related $\SZ^e$-basis for the ultra-multiplets, which accommodates both ultra-multiplets and their twisted variants, extending the representation\eqs{UM1}{UM2} and its Klein-flip.

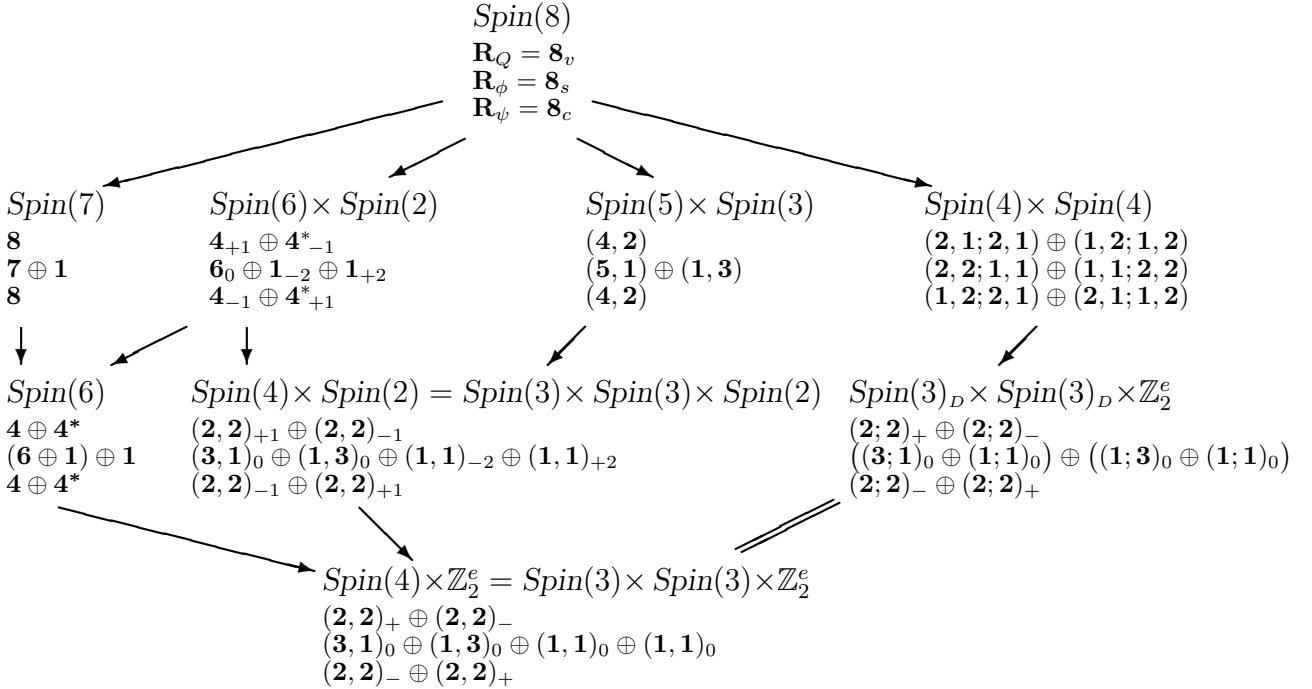
\begin{figure}[ht]
 \begin{center}
  \begin{picture}(170,90)(4,0)
   \put(80,85){\hbox to0pt{\hss\parbox[t]{30mm}{
      $\Spin(8)$\footnotesize\\[0mm]
      ${\bf R}_Q={\bf8}_v$\\[-1mm]
      ${\bf R}_\f={\bf8}_s$\\[-1mm]
      ${\bf R}_\j={\bf8}_c$}\hss}}
   \put(18,60){\hbox to0pt{\hss\parbox[t]{30mm}{
      $\Spin(7)$\footnotesize\\[0mm]
      ${\bf8}$\\[-1mm]
      ${\bf7}\oplus{\bf1}$\\[-1mm]
      ${\bf8}$}\hss}}
   \put(50,60){\hbox to0pt{\hss\parbox[t]{40mm}{
      $\Spin(6){\times}\Spin(2)$\footnotesize\\[0mm]
      ${\bf4}_{+1}\oplus{\bf4^*}\!_{-1}$\\[-1mm]
      ${\bf6}_0\oplus{\bf1}_{-2}\oplus{\bf1}_{+2}$\\[-1mm]
      ${\bf4}_{-1}\oplus{\bf4^*}\!_{+1}$}\hss}}
   \put(95,60){\hbox to0pt{\hss\parbox[t]{30mm}{
      $\Spin(5){\times}\Spin(3)$\footnotesize\\[0mm]
      $({\bf4},{\bf2})$\\[-1mm]
      $({\bf5},{\bf1})\oplus({\bf1},{\bf3})$\\[-1mm]
      $({\bf4},{\bf2})$}\hss}}
   \put(150,60){\hbox to0pt{\hss\parbox[t]{50mm}{
      $\Spin(4){\times}\Spin(4)$\footnotesize\\[0mm]
      $({\bf2,1;2,1})\oplus({\bf1,2;1,2})$\\[-1mm]
      $({\bf2,2;1,1})\oplus({\bf1,1;2,2})$\\[-1mm]
      $({\bf1,2;2,1})\oplus({\bf2,1;1,2})$}\hss}}
   \put(18,35){\hbox to0pt{\hss\parbox[t]{30mm}{
      $\Spin(6)$\footnotesize\\[0mm]
      ${\bf4}\oplus\bs{4^*}$\\[-1mm]
      $({\bf6}\oplus{\bf1})\oplus{\bf1}$\\[-1mm]
      ${\bf4}\oplus\bs{4^*}$}\hss}}
   \put(70,35){\hbox to0pt{\hss\parbox[t]{85mm}{
      $\Spin(4){\times}\Spin(2)
       =\Spin(3){\times}\Spin(3){\times}\Spin(2)$\footnotesize\\[0mm]
      $({\bf2,2})_{+1}\oplus({\bf2,2})_{-1}$\\[-1mm]
      $({\bf3,1})_0\oplus({\bf1,3})_0\oplus({\bf1,1})_{-2}\oplus({\bf1,1})_{+2}$\\[-1mm]
      $({\bf2,2})_{-1}\oplus({\bf2,2})_{+1}$}\hss}}
   \put(150,35){\hbox to0pt{\hss\parbox[t]{70mm}{
      $\Spin(3)_{\sss D}{\times}\Spin(3)_{\sss D}{\times}\ZZ_2^e$\footnotesize\\[0mm]
      $({\bf2;2})_+\oplus({\bf2;2})_-$\\[-1mm]
      $\big(({\bf3;1})_0\oplus({\bf1;1})_0\big)\oplus
       \big(({\bf1;3})_0\oplus({\bf1;1})_0\big)$\\[-1mm]
      $({\bf2;2})_-\oplus({\bf2;2})_+$}\hss}}
   \put(90,10){\hbox to0pt{\hss\parbox[t]{90mm}{
      $\SZ^e=\Spin(3){\times}\Spin(3){\times}\ZZ_2^e$
       \footnotesize\\[0mm]
      $({\bf2,2})_+\oplus({\bf2,2})_-$\\[-1mm]
      $({\bf3,1})_0\oplus({\bf1,3})_0\oplus
       ({\bf1,1})_0\oplus({\bf1,1})_0$\\[-1mm]
      $({\bf2,2})_-\oplus({\bf2,2})_+$}\hss}}
   \put(61,75){\vector(-4,-1){45}}
   \put(81,75){\vector(4,-1){45}}
   \put(64,70){\vector(-2,-1){10}}
   \put(79,70){\vector(2,-1){10}}
   \put(5,45){\vector(0,-1){5}}
   \put(27,45){\vector(-2,-1){10}}
   \put(35,45){\vector(0,-1){5}}
   \put(80,45){\vector(-1,-1){5}}
   \put(140,45){\vector(-1,-1){5}}
   \put(10,21){\vector(4,-1){34}}
   \put(50,21){\vector(1,-1){7}}
   \put(113,22){\line(-2,-1){13}}
   \put(113.8,21.6){\line(-2,-1){13}}
  \end{picture}
 \end{center}
 \caption{\small The relevant subgroup chains of $\Spin(8)$. Throughout, we use: $\Spin(6){=}\SU(4)$, $\Spin(5){=}\Sp(4)$, $\Spin(4){=}\Spin(3)^2$, $\Spin(3){=}\SU(2)$, $\Spin(2){=}\textsl{U}(1)$ and $\Spin(1){=}\ZZ_2$. In the $\Spin(6){=}\SU(4)\to\Spin(4)$ projection, both spinors, ${\bf4}$ and $\bs{4^*}$, of $\Spin(6)$ map to the real 4-vector of $\Spin(4)$. $\Spin(3)_{\sss D}$ is the diagonal subgroup of $\Spin(3){\times}\Spin(3){=}\Spin(4)$. Finally, it is only the $\ZZ_2$ structure in $\ZZ_2^e$ that is a proper symmetry group; nevertheless, distinguishing between the two spinors of $\Spin(2)$ turns out possible and useful.}
 \label{f:SO8}
\end{figure}

\paragraph{Kinship Symmetry:}
A survey of the $\Gf$ for the various Adinkras, \ie, supermultiplets in the ultra-multiplet family, we may define the {\em\/maximal common symmetry\/} for the whole family. As the relevant subgroup-chains of $\Spin(8)$ presented in Figure~\ref{f:SO8}  %
show, this ``common denominator'' symmetry is $\SZ^e$, manifest in the computational framework set up by Ref.\cite{rGR0}. This fact explains why this one framework can indeed be used to describe each supermultiplet in the family. 

We are now also in position to resolve the apparent discrepancy between the degeneracies listed in Table~\ref{t:I} and the different counting one can obtain by inspecting the Adinkras in Table~\ref{t:E8}. For example, as mentioned above, all eight bosons in the $(8|8)$-dimensional Valise supermultiplet\eq{eO8} appear equivalent, which is emphasized by identifying them as spanning the {\em\/irreducible\/} representation ${\bf8}_s$ of $\Spin(8)$. Thus, the operation of raising any particular one of them is equivalent to raising any other one. Similarly, the seven ``un-raised'' bosons in the $(7|8|1)$-dimensional supermultiplet, with one bosonic node already raised, are also equivalent. Raising any particular one of them is equivalent to raising any other. This way of counting implies that there exist $\binom8r$ distinct although equivalent $(8{-}r|8|r)$-dimensional supermultiplets, obtained by raising $r$ bosonic (open) nodes from the Valise formation of the ultra-multiplet\eqs{eAB88}{eO8}. This degeneracy|$\binom8r$ distinct but equivalent $(8{-}r|8|r)$-dimensional supermultiplets|is much larger than the one observed in Table~\ref{t:I}.

The resolution is in the fact that $\Spin(8)$ contains a {\em\/continuum\/} of $\Spin(4)$ subgroups all of which are isomorphic by conjugation, and the analogous holds for all other subgroups. The computational framework of Ref.\cite{rGR0}, showcased in Section~\ref{s:UM}, fixes a particular $\Spin(4)\subset\Spin(8)$ subgroup. This significantly limits the distinct albeit equivalent node-raising options. After all, consider the fact that when two nodes are raised, we must identify these raised nodes with $A\to(\ddt^{-1}A)$ and $B\to(\ddt^{-1}B)$. This leaves the two triplets, $A_{\sss\ha\hb}$ and $B_{\sss\ha\hb}$ waiting to be raised. These being triplets of $\SZ$, raising any one node, or two, at that point is impossible within this basis without in fact raising either of the two entire triplets. On the other hand, and within the same basis, we {\em\/can\/} describe the result of three nodes having been raised, by, say, $A_{\sss\ha\hb}\to(\ddt^{-1}A_{\sss\ha\hb})$. Thus, although raising two nodes $(8|8)\to(6|8|2)$ and raising three nodes $(8|8)\to(5|8|3)$ are both perfectly describable, it is not possible to describe the transition $(6|8|2)\to(5|8|3)$, marked by ``$?$'':
\begin{equation}
  \vC{\begin{picture}(80,15)(0,-2)
       \put(0,5){$(A,B,A_{{}_{\ha\hb}},B_{{}_{\ha\hb}}|
                    \j_{{}_{\hK\ha}})$}
       \put(47,0){$(A_{{}_{\ha\hb}},B_{{}_{\ha\hb}}|
                   \j_{{}_{\hK\ha}}|A,B)$}
       \put(47,10){$(A,B,B_{{}_{\ha\hb}}|
                   \j_{{}_{\hK\ha}}|A_{{}_{\ha\hb}})$}
       \thinlines
        \put(37,5.5){\vector(3,-1){9}}
         \put(40,8.5){$\SSS\Chekk$}
        \put(37,6.5){\vector(3,1){9}}
         \put(42,4){$\SSS\Chekk$}
        \put(56,3.5){\vector(0,1){5}}
         \put(56.5,5){\scriptsize\bf?}
      \end{picture}}
\end{equation}
without changing the $\SZ^e$-basis in the process.

This proves that there exist many distinct but equivalent $\SZ^e$-bases, and that more than one may be needed when describing not just any one of the supermultiplets in the family, but also the operation of changing one into another by means of the node-raising operation.

\paragraph{A Complementary $\SZ^e$-Basis:}
Although the definition~\ref{dUM} demonstrates the existence of both an ultra-multiplet and a twisted ultra-multiplet, the explicit construction\eqs{UM1}{UM2} affords encoding both variants only through a Klein-flip, as discussed above.

However, we can use the triality of $\Spin(8)$ and embed $\SZ\subset\Spin(8)$ so that
\begin{subequations}
 \label{MU1}
 \begin{alignat}{3}
 {\bf R}_\rD&=({\bf1,1})_0\oplus{{\bf1,1}}_0\oplus({\bf3,1})_0\oplus({\bf1,3})_0
 \quad&:\qquad
 &\rD_\pm \oplus \rD_\pm^{\sss\ha\hb},\\
 {\bf R}_\f&=({\bf2,2})_+\oplus({\bf2,2})_-
 \quad&:\qquad
 &\f_{\sss\hK\,\ha}:=\bs{\F}_{\sss\hK\,\ha}|,\\
 {\bf R}_\j&=({\bf2,2})_+\oplus({\bf2,2})_-
 \quad&:\qquad
 &\j_{\sss\hK\,\ha}:=\bs{\J}_{\sss\hK\,\ha}|,
\end{alignat}
\end{subequations}
where
 $\rD_\pm^{\sss\ha\hb}
 =\pm\inv2\ve^{\sss\ha\hb}{}_{\sss\hg\hd}\,\rD_\pm^{\sss\hg\hd}$, so that
 $\rD_\pm^{\sss\ha\hb}\d_{\sss\hb\hg}\rD_\pm^{\sss\hg\hd}
 =-\|\rD_\pm^{^{_{**}}}\|^2\d^{\ha\hd}$ and
 $\rD_+^{\sss\ha\hb}\d_{\sss\hb\hg}\rD_-^{\sss\hg\hd}
 =\rD_-^{\sss\ha\hb}\d_{\sss\hb\hg}\rD_+^{\sss\hg\hd}$,
and $\rD_\pm$ are labeled so as to accompany $\rD_\pm^{\sss\ha\hb}$. With these, it is straightforward to prove that the system of super-differential relationships
\begin{subequations}
 \label{MU2}
 \begin{alignat}{3}
 \rD_+\bs{\F}_{\sss\hK\,\ha}
  &=i\,\bs{\J}_{\sss\hK\,\ha},
  &\qquad
 \rD_+\bs{\J}_{\sss\hK\,\ha}
  &=(\ddt\bs{\F}_{\sss\hK\,\ha}),\\
 \rD_+^{\sss\ha\hb}\bs{\F}_{\sss\hK\,\hg}
  &=i\,(\s^3)_{\sss\hK}{}^{\sss\hL}
      \D_+^{\sss\ha\hb}{}_{\sss\hg}{}^{\sss\hd}\,\bs{\J}_{\sss\hL\,\hd},
  &\qquad
 \rD_+^{\sss\ha\hb}\bs{\J}_{\sss\hK\,\hg}
  &=-(\s^3)_{\sss\hK}{}^{\sss\hL}
      \D_+^{\sss\ha\hb}{}_{\sss\hg}{}^{\sss\hd}\,(\ddt\bs{\F}_{\sss\hL\,\hd}),\\
 \rD_-\bs{\F}_{\sss\hK\,\ha}
  &=\pm i\,\ve_{\sss\hK}{}^{\sss\hL}\bs{\J}_{\sss\hL\,\ha},
  &\qquad
 \rD_-\bs{\J}_{\sss\hK\,\ha}
  &=\mp\ve_{\sss\hK}{}^{\sss\hL}\,(\ddt\bs{\F}_{\sss\hL\,\ha}),\label{UTU}\\
 \rD_-^{\sss\ha\hb}\bs{\F}_{\sss\hK\,\hg}
  &=i\,(\s^1)_{\sss\hK}{}^{\sss\hL}
      \D_-^{\sss\ha\hb}{}_{\sss\hg}{}^{\sss\hd}\,\bs{\J}_{\sss\hL\,\hd},
  &\qquad
 \rD_-^{\sss\ha\hb}\bs{\J}_{\sss\hK\,\hg}
  &=-(\s^1)_{\sss\hK}{}^{\sss\hL}
      \D_-^{\sss\ha\hb}{}_{\sss\hg}{}^{\sss\hd}\,(\ddt\bs{\F}_{\sss\hL\,\hd}),\\
  \D_\pm^{\sss\ha\hb}{}_{\sss\hg}{}^{\sss\hd}
  &:=2\d^{\sss[\ha}_{\>\sss\hg}\d^{\sss\hb]\hd}
     \pm\ve^{\sss\ha\hb}{}_{\sss\hg}{}^{\sss\hd},\quad&
  \text{so that}\quad\inv2\ve^{\sss\ha\hb}{}_{\sss\hat\e\hat\vf}\,
        \D_\pm^{\sss\hat\e\hat\vf}{}_{\sss\hg}{}^{\sss\hd}
  &=\pm\D_\pm^{\sss\ha\hb}{}_{\sss\hg}{}^{\sss\hd},
\end{alignat}
\end{subequations}
defines an $(8|8)$-dimensional supermultiplet of $N=8$-extended supersymmetry, just as do the equations\eqs{UM1}{UM2}. The choice of the upper/lower sign in the relations\eq{UTU} provides precisely the twist between the ultra-multiplet and the twisted ultra-multiplet. In retrospect, it should be noted that $\rD_+,\rD_+^{\sss\ha\hb},\rD_-^{\sss\ha\hb},\rD_-$ may be corresponded to the 0-form, self-dual 2-form, anti-self-dual 2-form and 4-form of $\Spin(4)$. Somewhat akin to the degeneracies in Table~\ref{t:I}, we can change signs of these four operators in various combinations, providing a total of $2^4=16$ sign-choices|precisely as within the system\eq{PE8}.

We thus conclude that the $\SZ^e$-basis\eqs{MU1}{MU2} captures the inequivalent sign-choices in the quasi-projector system\eq{PE8}. On the other hand, since both the bosons and the fermions are now ``packaged'' as pairs of 4-plets, it is not possible to node-raise the bosons nor node-lower the fermions one-by-one, but only node-raise or lower all eight.
 In turn, the $\SZ^e$-basis\eqs{UM1}{UM2} is well suited to discuss the incremental node-raising operations through ultra-multiplet family, but presents a way to twist between the ultra-multiplet and its twisted variant only through a Klein-flip.

\paragraph{Generic Background:}
Although $\Spin(8)$ and its unique triality seems to play a prominent role in the present analysis, we now trace the existence of this computationally useful $\SZ^e$-basis to a generic feature of $\Spin(2n)$ groups, and in fact supermultiplets.

Every $\Spin(2n)$ group has two minimal spinor representations, ${\bf R}_s$ and ${\bf R}_c$ and a regular $\SU(n)\times\sU(1)$ subgroup, unambiguously defined by the decompositions:
\begin{equation}
 \begin{array}{@{}r@{\,:~}r@{\,}lr@{\,}lr@{\,}l@{}}
   \Spin(2n)
    & {\bf R}_v&=\IR^{2n}
    & {\bf R}_s&\approx\IR^{2^{n-1}}
    & {\bf R}_c&\approx\IR^{2^{n-1}}\\[1mm]
   \SU(n)\times\sU(1)
    & {\bf R}_v&\to\IC^n_{+1}+(\IC^n)^*_{-1}
    & {\bf R}_s&\to\bigoplus_{p\text{ even}}(\wedge^p\IC^n)_{q_p}
    &{\bf R}_c&\to\bigoplus_{p\text{ odd}}\wedge^p\IC^n_{q_p}
 \end{array}
 \label{eSpin(2n)}
\end{equation}
where ${\bf R}_v$ is the standard $2n$-vector representation, and the $\sU(1)$ charges $q_p$ are specified in the Appendix~\ref{a:Spin}.
 In the first row of\eq{eSpin(2n)}, we ``forget'' that ${\bf R}_s,{\bf R}_c$ admit a complex structure and are each other's complex conjugate for $n=1\pmod2$, and admit a real {\em\/vs\/}.\ pseudo-real structure when $n=0\pmod4$ {\em\/vs\/}.\ $n=2\pmod4$, respectively. Also, $\wedge^p\IC^n$ denotes the vector space of complex $p$-forms in complex $n$-dimensional space.

For $\Spin(2n{+}1)$, the analogous regular subgroup is defined by
\begin{equation}
 \begin{array}{@{}r@{\,:~}r@{\,}lr@{\,}l@{}}
   \Spin(2n{+}1)
    & {\bf R}_v&=\IR^{2n+1}
    & {\bf R}_s&\approx\IR^{2^n}\\[1mm]
   \SU(n)\times\sU(1)
    & {\bf R}_v&\to\IC^n_{+1}+(\IC^n)^*_{-1}+\IR^1_0
    & {\bf R}_s&\to\bigoplus_{p=0}^n(\wedge^p\IC^n)_{q_p}
 \end{array}
 \label{eSpin(2n+1)}
\end{equation}
where the $\sU(1)$ charges are determined by the embedding $\Spin(n){\times}\sU(1)\subset\Spin(2n)\subset\Spin(2n{+}1)$, as detailed in Appendix~\ref{a:Spin} and differ from the conventions of Ref.\cite{rSsky}.

The special unitary subgroup $\SU(n)$ itself has a special real subgroup, $\Spin(n)\subset\SU(n)$, defined by turning the ground field real and including the invariant positive-definite metric, the Kronecker $\d$-symbol by choice of basis. This however permits, for even $n$, to further decompose the (now real-valued!) middle-forms $\wedge^{n/2}\IR^n$ in\eq{eSpin(2n)} into the self-dual and the anti-self-dual halves.

For the case at hand, we have the subgroup chain (see Appendix~\ref{a:Spin} for details):
\begin{subequations}
 \label{eSpin824}
\begin{alignat}{5}
  \Spin(8) &~&\supset~& \SU(4)\times\sU(1)
           &~&\supset \Spin(4)\times\ZZ_2=\SU(2)^2\times\ZZ_2\\[1mm]
  {\bf R}_Q={\bf8}_v  &&\to~& {\bf4}_{+1}\oplus\bs{4^*}\!_{-1}
                      &&\to~  ({\bf2,2})_+\oplus({\bf2,2})_-;  \label{BQ}\\
  {\bf R}_\f={\bf8}_s  &&\to~& {\bf1}_{-2}\oplus{\bf6}_0\oplus{\bf1}_{+2}
                      &&\to~  ({\bf1,1})_0\oplus ({\bf3,1})_0
                               \oplus({\bf1,3})_0\oplus ({\bf1,1})_0; \label{Bf}\\
  {\bf R}_\j={\bf8}_c &&\to~& {\bf4}_{-1}\oplus\bs{4^*}\!_{+1}
                      &&\to~  ({\bf2,2})_-\oplus({\bf2,2})_+. \label{Bj}
\end{alignat}
\end{subequations}
The $\sU(1)$-charges\cite{rSsky} are seen to agree with the values of the $\hK$-type indices without modification on the fermionic superfields and component fields and on $\rD_{\sss\hK\,\ha}$ and $Q_{\sss\hK\,\ha}$, but only upon a (mod)~2 reduction to ``0'' in the middle row\eq{Bf}, indicating the absence of the $\hK$-type indices.

From the above analysis of the node-raising dependence of $\Gf$, it is evident that for each chromotopology, it is the Valise Adinkra which offers the maximal $\Gf$.
 Guided by the $\Gf$-assignment\eq{eO8}, we assign the two irreducible spinor representations of a suitable $\Spin$-group to the bosons and fermions, respectively:
\begin{equation}
  \Span(\f_1,\cdots,\f_m) = {\bf R}_s,\quad
  \Span(\j_1,\cdots,\j_m) = {\bf R}_c,\quad
   \text{of}\quad \Gf=\Spin(2N),\quad
   \text{for all }N.
 \label{eSpin(2N)}
\end{equation}
This identification is made precise in a formal Fock-space construction:
\begin{subequations}
 \label{exp}
 \begin{alignat}{7}
 \textsl{\small\bfseries\boldmath Spin$(2N)$}~\text{\small\Bf Root Lattice}
  &&&\text{\small\Bf\boldmath$Q$-monomials}
  &&\hbox to10pt{\small\Bf comp.\ fields\hss}\nn\\*
 \he_\rI&\longmapsto&~
     &Q^{\he_\rI}:=Q_1^0\cdots Q_{I-1}^0\,Q_\rI^1\,Q_{I+1}^0\cdots Q_N^0
                                    &\,=\,&Q_\rI,\\
 \he_\rI+\he_\rJ&\mapsto&~
     &Q^{\he_\rI+\he_\rJ}:=Q_1^0\cdots Q_\rI^1\cdots Q_\rJ^1\cdots Q_N^0
                                    &\,=\,&Q_\rI Q_\rJ, \label{eQIQJ}\\
 (-\inv2,-\inv2,-\inv2,-\inv2,\cdots,-\inv2)&\mapsto&
     &Q^{(0,0,0,0,\cdots,0)}\ket{0} &\,=\,&\ket{0}&&=:\f_0,\\
 (+\inv2,-\inv2,-\inv2,-\inv2,\cdots,-\inv2)&\mapsto&
     &Q^{(1,0,0,0,\cdots,0)}\ket{0} &\,=\,&Q_1\ket{0}&&=:\j_1,\\
 (+\inv2,+\inv2,-\inv2,-\inv2,\cdots,-\inv2)&\mapsto&
     &Q^{(1,1,0,0,\cdots,0)}\ket{0} &\,=\,&Q_1Q_2\ket{0}&&=:\f_{[12]},\\
     \textit{etc.}&&&\textit{etc.}&&\textit{etc.}\nn
\end{alignat}
\end{subequations}
where the $Q$'s in the $Q$-monomials\eq{exp} are always ordered lexicographically, say, resolving the ambiguity stemming from the fact that the addition of root vectors is commutative, whereas the product of different $Q_\rI$'s is anticommutative. In this construction,
\begin{equation}
  {\bf R}_v = \Span\big((\pm1,0,0,\cdots,0),(0,\pm1,0,\cdots,0),
                         \cdots,(0,\cdots,0,\pm1)\big)
 \label{eV}
\end{equation}
maps ${\bf R}_s\iff{\bf R}_c$ and so must contain the $Q$'s. Note however, that this
distinguishes between the $\ddt$-less from the $\ddt$-action of the $Q$'s, as in
\begin{equation}
 \left.\begin{aligned}
  Q_1 \f_0 &= \j_1,\\
  Q_1 \f_{[23]}&=\j_{[123]},\\
     &\mkern-1mu\textit{etc.}
 \end{aligned}\right\}
  \qquad\textit{vs.}\qquad
 \left\{\begin{aligned}
  Q_1 \j_1 &= i(\ddt\f_0),\\
  Q_2 \f_{[23]}&=i(\ddt\j_3),\\
    &\mkern-1mu \textit{etc.}
 \end{aligned}\right.
\end{equation}
Given the familiar superspace realizations:
\begin{equation}
  Q_\rI=i\vd_\rI+\d_{\rI\rJ}\,\q^\rJ\,\ddt,\qquad
  \rD_\rI=\vd_\rI+i\d_{\rI\rJ}\,\q^\rJ\,\ddt,
 \label{eSSQD}
\end{equation}
it follows that
\begin{equation}
  (0,{\cdots},0,+1,0,{\cdots},0) ~\mapsto~ \inv2(\rD_\rI-iQ_\rI),\qquad
  (0,{\cdots},0,-1,0,{\cdots},0) ~\mapsto~ \inv2(\rD_\rI+iQ_\rI),
 \label{eVQD}
\end{equation}
where the nonzero entries are in the $\rI^\text{th}$ position.
 Therefore, for all $N\in\IN$, each $N$-cubical, {\em\/a priori\/} unconstrained and unprojected Valise supermultiplet\eq{eSM=SF} admits a $\Gf=\Spin(2N)$ action specified by\eq{eSpin(2N)} and\eqs{eV}{eVQD}. As nodes are raised, $\Gf$ changes through the subgroup chains of $\Spin(2N)$, not dissimilar to the discussion in Section~\ref{s:SymUM}. Theorems~5.3 and~7.6 of Ref.\cite{r6-1} prove that all supermultiplets with the same chromotopology, however variously ``hung''\Ft{A particular ``hanging'' of a supermultiplet specifies one of the consistent assignments of the component fields' engineering units. The term alludes to fixing the components with locally (within the network of connections defined by supersymmetry transformation) maximal engineering units at corresponding heights and letting the supermultiplet hang freely from these, akin to a hanging garden or a macram\'e\cite{r6-1}.}, can be obtained from the Valise\eq{eSM=SF}, as can their superfield representations.

In turn, when a supermultiplet is projected by a $\ZZ_2$ reflection corresponding to an operator of the type
(\ref{PE8}), $\Gf$ reduces in rank by one: As discussed in Section~\ref{s:E8}, each such multiplet is annihilated by some projection operator such as $\hat\P^-_{1234}$, implying that $Q_1Q_2Q_3Q_4\simeq+H^2$ when acting on this supermultiplet; equivalently, $\rD_1\rD_2\rD_3\rD_4\simeq+H^2$ on the superfield realization\eq{eSM=SF}. This permits expressing one $Q$ and one $D$ in terms of the others, and so reduces $\dim({\bf R}_v)$ by two, and in turn, $\Spin(2N)\to\Spin(2N{-}2)$. Correspondingly, as evident from the exponential mapping\eq{exp}, the number of component (super)fields in the Valise\eq{eSM=SF} reduces by a factor of two: $m=2^{N-1}\to2^{N-2}$. After $k$ such projections, $\Spin(2N)\to\Spin(2\6(N{-}k))$ and $m=2^{N-1}\to2^{N-k-1}$.

The maximum number of such projections is $\inv2N$, and can be achieved only for $N=0\pmod8$\cite{r6-3}; in general, the number of projections is limited by:
\begin{equation}
 \vk(N):=
  \begin{cases}
   0 &\text{for $0\leq N<4$};\\
   \big\lfloor\frac{(N-4)^2}{4}\big\rfloor+1 &\text{for $N=4,5,6,7$};\\
   \vk(N{-}8)+4 &\text{for $N>7$, recursively}.
  \end{cases}
 \label{eKmax}
\end{equation}
which is closely related to the Radon-Hurwitz function\cite{rPT}.
 In such cases $\Gf=\Spin(2N)\to\Spin(N)$ and $m=2^{N-1}\to2^{N/2-1}$|precisely the case for $N=8$ ultra-multiplet, being the case-study in this paper. For these values of $N$, the minimal supermultiplet|maximally projected from the one with $N$-cubical chromotopology|is most compact.
 It is fascinating that precisely in these $N=0\pmod8$ cases, the doubly-even binary linear block codes offer error-detecting and error-correcting encryption with minimal information-theoretic Shannon entropy.

Finally, the general existence of the $\Spin(n){\times}\SU(1)\subset\Spin(2n)$ subgroup implies the existence of a $\Spin(n){\times}\ZZ_2^e$-basis for all $n$. However, its utility in computational effectiveness peaks for $2n=N\leq8$, in the sense that this basis permits a unified description of all supermultiplets within same family, \ie, all supermultiplets with the same chromotopology for the same $N\leq8$ extended supersymmetry. For $N>8$, this utility diminishes; see Appendix~\ref{a:Spin}.

\section{The $N=8$  Super-Zeemann Effect Multiplex}
 \label{s:SZEM}
The work of Ref.\cite{r6-7a} introduced a class of models with a coupling of background magnetic fluxes to worldline models with arbitrarily $N$-extended supersymmetry.  However, the formulation presented there relied solely on the $N = 1$ superfield formulation.  Since we are concentrating on manifestly $N = 8$ supermultiplet formulations\cite{rGLP,r6-2} of the ultra-multiplet complex in the current work, we have the opportunity to re-visit the previous work specifically for ultra-multiplets.

Following the approach of Ref.\cite{r6-7a}, we begin by introducing $M$ pairs of ultra-multiplets.  The component fields of both members of a pair of $\vec a = 0$
ultra-multiplets can be denoted by
\begin{equation}
  \big(\, A^p,~ B^p,~ A_{\sss\ha\hg}^p\,,~
          B_{\sss\Hat\a\Hat\g}^p\,\big|\>
          \j\1_{\sss{\Hat\rK\,\ha}}^{\,p}\,\big)
  \qquad\text{and}\qquad
  \big(\, \Tw{A}^p,~ \Tw{B}^p,~ \Tw{A}_{\sss\Hat\a\Hat\g}^p\,,~
           \Tw{B}_{\sss\Hat\a\Hat\g}^p\,\big|\>
            \Tw{\j}_{\sss\Hat\rK\,\ha}^{\,p}   \, \big)
\label{ultraSZEM1}
\ee
where the indices $p,q$ take on values $1,\dots,M$, counting the pairs\eq{ultraSZEM1}.  Next we introduce a constant 2-form of background fluxes denoted by
 ${\cal F}_{pq}$ following the prescription 
given in Ref.\cite{r6-7a} and write:
\begin{equation}
 {\cal L}_\text{Flux}
  ={\cal F}_{pq}\,
     \Big[\,A^p(\ddt\Tw{A}^q)
         + B^p(\ddt\Tw{B}^q)
         + \fracm14\,A^{{\sss\Hat\a\Hat\g}\,p}(\ddt\Tw{A}_{\sss\Hat\a\Hat\g}^q)
         + \fracm14\,B^{{\sss\Hat\a\Hat\g}\,p}(\ddt\Tw{B}_{\sss\Hat\a\Hat\g}^q)
         + i\,\j^{{\sss\Hat\rK\,\ha}\,p}\,\Tw{\j}_{\sss\Hat\rK\,\ha}^q\,\Big],
 \label{ultraFLUX}
\end{equation}
where the supersymmetry invariance of this action demands that both ultra-multiplets
be as defined in \Eqs{UM1}{UM2}. Analogous Lagrangians using the representation\eqs{MU1}{MU2} is as straightforward.

In the process of constructing the coupling of the $\vec a $ = 0 ultra-multiplets to
magnetic fluxes as described by (\ref{ultraFLUX}) we also have found it is possible 
to introduce pure mass terms and mixed mass-flux terms for ultra-multiplet pairs
with $\vec a$ $\ne$ 0 as well.  This circumstance owes to the existence of a
superinvariant of the form
\begin{equation}
 {\cal L}_\text{Pair}
  =m\,\d_{pq}\,
     \Big[\,A^p(\ddt\Tw{A}^q)
         + B^p(\ddt\Tw{B}^q)
         + \fracm14\,A^{{\sss\Hat\a\Hat\g}\,p}(\ddt\Tw{A}_{\sss\Hat\a\Hat\g}^q)
         + \fracm14\,B^{{\sss\Hat\a\Hat\g}\,p}(\ddt\Tw{B}_{\sss\Hat\a\Hat\g}^q)\\
         + i\,\j^{{\sss\Hat\rK\,\ha}\,p}\,\Tw{\j}_{\sss\Hat\rK\,\ha}^q\,\Big].
 \label{ultraPair}
\end{equation}
for the $\vec a$ = 0 case. Evidently, the Lagrangian\eq{ultraPair} is a special case of\eq{ultraFLUX}, whereupon the mass of all these fields, as introduced in\eq{ultraPair}, may be regarded as induced and controlled {\em\/via\/} the coupling to a background flux.

By performing node lifts\eq{UM5} on this expression a wide variety of bilinear actions may be constructed for all choices of $\vec a$ listed in Table~\ref{t:I}, one for every member of the family of supermultiplets\eqs{eO8}{eAF88}. Since the alternate representation of the ultra-multiplet and its twisted variant\eqs{MU1}{MU2} does not admit individual bosonic node raises without spoiling the underlying $\SZ^e$-basis, the task of constructing the most general model involving the ultra-multiplet, the twisted ultra-multiplet and their various node-raised versions seems to necessitate dropping the $\SZ^e$ basis and using instead the ``plain'' basis of\eq{eSM=SF}.

\section{Conclusions and Outlook}
 \label{s:CO}
In this note, we have examined the family of minimal, $8{+}8$-dimensional off-shell representations of worldline $N=8$ extended supersymmetry; jointly furnishing an example of the ``root superfield'' formalism of Ref.\cite{rGLP}. These off-shell supermultiplets are faithfully depicted by the Adinkras\eqs{eO8}{eAF88}\cite{rA,r6-1}, and admit effective symmetry groups that depend on the hanging: the particular assignment of the component fields' engineering units; see Table~\ref{t:E8}. These groups of effective symmetries are all subgroups of the maximal one, $\Spin(8)$|exhibited by the Valise supermultiplet\eq{eO8} and\eq{eAF88}. In turn, these groups of effective symmetries all have a common subgroup, $\SZ$, and this is the underlying symmetry group of the particular basis used in the original construction of Ref.\cite{rGR0}, which also features a formally non-associative ``extension'' $\ZZ_2\to\ZZ_2^e$.

The supermultiplets in the ultra-multiplet family all have the $E_8$ chromotopology\cite{r6-3,r6-3.2}, so their Adinkras\eqs{eO8}{eAF88} have the structure of the $(\ZZ_2)^4$-quotient of the 8-cube encoded by the doubly-even binary linear block code $e_8$. This induces a direct correspondence between the so-called even/odd equivalence classes of $E_8$ root lattices and the (un)twisted variants of the ultra-multiplets\eqs{MU1}{MU2}; see Appendix~\ref{a:E8}.

Along the way, we present the ultra-multiplet in terms of a super-differentially constrained system of off-shell superfields\eqs{UM1}{UM2} and\eqs{MU1}{MU2}. The end of Appendix~\ref{a:D2Q} provides a proof that such a constrained system of superfields may be corresponded to each of the trillions of inequivalent Adinkras found through the classification efforts of Refs.\cite{r6-1,r6-3,r6-3.2}. This then provides a {\em\/second\/} superfield representation of every Adinkraic worldline supermultiplet, complementing the combined construction of Refs.\cite{r6-1,r6-1.2}.

Section~\ref{s:SZEM} lists quadratic Lagrangians for the ultra-multiplet that provide the standard kinetic terms, but also terms that mix two ultra-multiplets depending on the interaction with a 2-form of external, background fluxes. It should be possible to extend the methods of Refs.\cite{rGLP,r6-2} so as to construct fully interactive, non-linear $\s$-models for the supermultiplets in the ultra-multiplet family, perhaps not unlike those presented for minimal supermultiplets of $N=4$ extended supersymmetry in Ref.\cite{rGHR}.

One of the most important messages we believe can be gleaned from our current study of the ultra-multiplet is its implications for the symmetries that can occur in off-shell versions of supersymmetric systems.

All of the multiplets discussed in this work provide realizations of $N = 8$ worldline supersymmetry. Importantly however, the symmetry groups under which the eight supercharges transform for most of them is {\em\/not\/} $\textsl{O}(8)$ as might be naively expected from the form of Eqs\eq{eSuSy} alone. In fact, depending upon which set of raising is performed, many distinct groups are found to provide the effective symmetry groups, $\Gf$, under which the supercharges transform.

One of the well-accepted tenets of conventional wisdom about supersymmetry representation theory is that, in the context of Poincar\'e supersymmetry in $d$ dimensions, the $N$ supercharges are ``bundled'' into $\cal N$ minimal spinors of $\Spin(1,d{-}1)$, and provide a representation of $\textsl{O}({\cal N})$ so-called $R$-symmetry. 

The present analysis of the ultra-multiplet suggests that this is not generally the case: On the $d=1$ worldline, the Lorentz symmetry reduces to $\Spin(1)=\ZZ_2$, its minimal spinors are 1-dimensional, and the manifest $\sO(8)$ symmetry of the supersymmetry algebra\eq{eSuSy} is indeed this $R$-symmetry of conventional wisdom. In all but the Valise supermultiplets\eq{eAB88} and\eq{eAF88}, this $\sO(8)$ is broken to its various appropriate subgroups, as detailed in Tables~\ref{t:E8} and~\ref{t:O8} and Figure~\ref{f:SO8}.

Granted, our examples are all restricted to the one-dimensional worldline, but if this qualitative behavior should persist in $d>1$ dimensional theories, it would provide a route by which to surmount the famous off-shell no-go theorem of Siegel and Ro\v{c}ek\cite{rRSnogo}, for $d=4$. In fact, one of the assumptions in the derivation of Ref.\cite{rRSnogo} was precisely the presence of an $\textsl{O}({\cal N})$ symmetry. In turn, these authors indicate, in an oft overlooked portion of that paper, that precisely the relaxing of this assumption offers a possibility to go off-shell. Finally, the relaxation of this assumption of maximal $R$-symmetry is also the key for the combinatorial explosion of supermultiplets proved in Refs.\cite{r6-3,r6-3.2}.

\vfill
\begin{flushright}\sl
 Symmetry is in the eyes of the beholder.\\[-1mm]
 $\sim$~Lieh-tzu\,\\[2mm]
\end{flushright}
 \vfill
\bigskip\paragraph{\bfseries Acknowledgments:}
We would like to thank C.F.~Doran, K.M.~Iga and G.D.~Landweber for helpful discussions without which this work could not have been completed.
 This research was supported in part by the endowment of the John S.~Toll Professorship, the University of Maryland Center for String \& Particle Theory, National Science Foundation Grant PHY-0354401, and Department of Energy Grant DE-FG02-94ER-40854. TH is a visiting professor at the Physics Department of the Faculty of Natural Sciences of the University of Novi Sad, Serbia, and wishes to thank for the recurring hospitality and resources. The Adinkras were drawn with the aid of the {\em Adinkramat\/}~\copyright\,2008 by G.~Landweber.

\clearpage\appendix
\section {Supersymmetry Transformation Rules}
 \label{a:D2Q}
Refs.\cite{r6-1,r6-3,r6-3.2,r6-3.4} classify some trillions of supermultiplets each of which that can be represented by an Adinkra such as\eq{eAB88} and which should be regarded merely as the ``simpler'' building blocks from which to construct myriads of additional supermultiplets by the usual technique of tensoring, symmetrizing and contracting.

Up to conventions, the $\rL_\rI$ matrices read off of the Adinkra\eq{eAB88} are the
same ones described in the work of Ref\cite{rGR1}:
\begin{equation}
  \begin{array}{@{} r|rrrrrrrr @{}}
    ¥ &\bs{\f_1} &\bs{\f_2} &\bs{\f_3} &\bs{\f_4}
      &\bs{\f_5} &\bs{\f_6} &\bs{\f_7} &\bs{\f_8}  \\ 
    \toprule
 \C1{\bs{\rL_1}} & \j_1 & \j_2 & \j_3 & \j_4 & \j_5 & \j_6 & \j_7 & \j_8 \\
 \C2{\bs{\rL_2}} &-\j_2 & \j_1 & \j_4 &-\j_3 & \j_6 &-\j_5 &-\j_8 & \j_7 \\
 \C3{\bs{\rL_3}} &-\j_3 &-\j_4 & \j_1 & \j_2 & \j_7 & \j_8 &-\j_5 &-\j_6 \\
 \C4{\bs{\rL_4}} &-\j_5 &-\j_6 &-\j_7 &-\j_8 & \j_1 & \j_2 & \j_3 & \j_4 \\
 \C5{\bs{\rL_5}} &-\j_4 & \j_3 &-\j_2 & \j_1 & \j_8 &-\j_7 & \j_6 &-\j_5 \\
 \C6{\bs{\rL_6}} &-\j_6 & \j_5 &-\j_8 & \j_7 &-\j_2 & \j_1 &-\j_4 & \j_3 \\
 \C7{\bs{\rL_7}} &-\j_7 & \j_8 & \j_5 &-\j_6 &-\j_3 & \j_4 & \j_1 &-\j_2 \\
 \C8{\bs{\rL_8}} &-\j_8 &-\j_7 & \j_6 & \j_5 &-\j_4 &-\j_3 & \j_2 & \j_1 \\
    \bottomrule
  \end{array}
 \label{eQ(phi)}
\end{equation}
or, alternatively:
\begin{subequations}
 \label{eLs}
\begin{align}
 \C1{\rL_1}=
 &\C1{\left[\begin{smallmatrix}
              1 & 0 & 0 & 0 & 0 & 0 & 0 & 0 \\
              0 & 1 & 0 & 0 & 0 & 0 & 0 & 0 \\
              0 & 0 & 1 & 0 & 0 & 0 & 0 & 0 \\
              0 & 0 & 0 & 1 & 0 & 0 & 0 & 0 \\
              0 & 0 & 0 & 0 & 1 & 0 & 0 & 0 \\
              0 & 0 & 0 & 0 & 0 & 1 & 0 & 0 \\
              0 & 0 & 0 & 0 & 0 & 0 & 1 & 0 \\
              0 & 0 & 0 & 0 & 0 & 0 & 0 & 1 \\
            \end{smallmatrix}\right]}&
 \C2{\rL_2}=
 &\C2{\left[\begin{smallmatrix}
              0 & -1 &  0 &  0 &  0 &  0 &  0 &  0 \\
              1 &  0 &  0 &  0 &  0 &  0 &  0 &  0 \\
              0 &  0 &  0 &  1 &  0 &  0 &  0 &  0 \\
              0 &  0 & -1 &  0 &  0 &  0 &  0 &  0 \\
              0 &  0 &  0 &  0 &  0 &  1 &  0 &  0 \\
              0 &  0 &  0 &  0 & -1 &  0 &  0 &  0 \\
              0 &  0 &  0 &  0 &  0 &  0 &  0 & -1 \\
              0 &  0 &  0 &  0 &  0 &  0 &  1 &  0 \\
            \end{smallmatrix}\right]}&
 \C3{\rL_3}=
 &\C3{\left[\begin{smallmatrix}
              0 &  0 & -1 &  0 &  0 &  0 &  0 &  0 \\
              0 &  0 &  0 & -1 &  0 &  0 &  0 &  0 \\
              1 &  0 &  0 &  0 &  0 &  0 &  0 &  0 \\
              0 &  1 &  0 &  0 &  0 &  0 &  0 &  0 \\
              0 &  0 &  0 &  0 &  0 &  0 &  1 &  0 \\
              0 &  0 &  0 &  0 &  0 &  0 &  0 &  1 \\
              0 &  0 &  0 &  0 & -1 &  0 &  0 &  0 \\
              0 &  0 &  0 &  0 &  0 & -1 &  0 &  0 \\
            \end{smallmatrix}\right]}\\
 \C4{\rL_4}=
 &\C4{\left[\begin{smallmatrix}
              0 &  0 &  0 &  0 & -1 &  0 &  0 &  0 \\
              0 &  0 &  0 &  0 &  0 & -1 &  0 &  0 \\
              0 &  0 &  0 &  0 &  0 &  0 & -1 &  0 \\
              0 &  0 &  0 &  0 &  0 &  0 &  0 & -1 \\
              1 &  0 &  0 &  0 &  0 &  0 &  0 &  0 \\
              0 &  1 &  0 &  0 &  0 &  0 &  0 &  0 \\
              0 &  0 &  1 &  0 &  0 &  0 &  0 &  0 \\
              0 &  0 &  0 &  1 &  0 &  0 &  0 &  0 \\
            \end{smallmatrix}\right]}&
 \C5{\rL_5}=
 &\C5{\left[\begin{smallmatrix}
              0 &  0 &  0 & -1 &  0 &  0 &  0 &  0 \\
              0 &  0 &  1 &  0 &  0 &  0 &  0 &  0 \\
              0 & -1 &  0 &  0 &  0 &  0 &  0 &  0 \\
              1 &  0 &  0 &  0 &  0 &  0 &  0 &  0 \\
              0 &  0 &  0 &  0 &  0 &  0 &  0 &  1 \\
              0 &  0 &  0 &  0 &  0 &  0 & -1 &  0 \\
              0 &  0 &  0 &  0 &  0 &  1 &  0 &  0 \\
              0 &  0 &  0 &  0 & -1 &  0 &  0 &  0 \\
            \end{smallmatrix}\right]}&
 \C6{\rL_6}=
 &\C6{\left[\begin{smallmatrix}
              0 &  0 &  0 &  0 &  0 & -1 &  0 &  0 \\
              0 &  0 &  0 &  0 &  1 &  0 &  0 &  0 \\
              0 &  0 &  0 &  0 &  0 &  0 &  0 & -1 \\
              0 &  0 &  0 &  0 &  0 &  0 &  1 &  0 \\
              0 & -1 &  0 &  0 &  0 &  0 &  0 &  0 \\
              1 &  0 &  0 &  0 &  0 &  0 &  0 &  0 \\
              0 &  0 &  0 & -1 &  0 &  0 &  0 &  0 \\
              0 &  0 &  1 &  0 &  0 &  0 &  0 &  0 \\
            \end{smallmatrix}\right]}\\
 \C7{\rL_7}=
 &\C7{\left[\begin{smallmatrix}
              0 &  0 &  0 &  0 &  0 &  0 & -1 &  0 \\
              0 &  0 &  0 &  0 &  0 &  0 &  0 &  1 \\
              0 &  0 &  0 &  0 &  1 &  0 &  0 &  0 \\
              0 &  0 &  0 &  0 &  0 & -1 &  0 &  0 \\
              0 &  0 & -1 &  0 &  0 &  0 &  0 &  0 \\
              0 &  0 &  0 &  1 &  0 &  0 &  0 &  0 \\
              1 &  0 &  0 &  0 &  0 &  0 &  0 &  0 \\
              0 & -1 &  0 &  0 &  0 &  0 &  0 &  0 \\
            \end{smallmatrix}\right]}&
 \C8{\rL_8}=
 &\C8{\left[\begin{smallmatrix}
              0 &  0 &  0 &  0 &  0 &  0 &  0 & -1 \\
              0 &  0 &  0 &  0 &  0 &  0 & -1 &  0 \\
              0 &  0 &  0 &  0 &  0 &  1 &  0 &  0 \\
              0 &  0 &  0 &  0 &  1 &  0 &  0 &  0 \\
              0 &  0 &  0 & -1 &  0 &  0 &  0 &  0 \\
              0 &  0 & -1 &  0 &  0 &  0 &  0 &  0 \\
              0 &  1 &  0 &  0 &  0 &  0 &  0 &  0 \\
              1 &  0 &  0 &  0 &  0 &  0 &  0 &  0 \\
            \end{smallmatrix}\right]}
\end{align}
\end{subequations}
or, finally, using
 $\Ione=\left[\begin{smallmatrix}1&0\\0&1\end{smallmatrix}\right]$,
 $\s^1=\left[\begin{smallmatrix}0&1\\1&0\end{smallmatrix}\right]$,
 $\s^2=\left[\begin{smallmatrix}0&-i\\i&0\end{smallmatrix}\right]$ so
 $(-i\s^2)=\left[\begin{smallmatrix}0&-1\\1&0\end{smallmatrix}\right]$, and
 $\s^3=\left[\begin{smallmatrix}1&0\\0&-1\end{smallmatrix}\right]$,
\begin{subequations}
 \label{eTensorLs}
\begin{alignat}{7}
 \C1{\rL_1}&=\C1{\Ione\otimes[\Ione\otimes\Ione]},
  \quad&\quad
 \C2{\rL_2}&=\C2{\s^3\otimes[\s^3\otimes(-i\s^2)]},\\
 \C3{\rL_3}&=\C3{\s^3\otimes[(-i\s^2)\otimes\Ione]},
  \quad&\quad
 \C4{\rL_4}&=\C4{(-i\s^2)\otimes[\Ione\otimes\Ione]},\\
 \C5{\rL_5}&=\C5{\s^3\otimes[\s^1\otimes(-i\s^2)]},
  \quad&\quad
 \C6{\rL_6}&=\C6{\s^1\otimes[\Ione\otimes(-i\s^2)]},\\
 \C7{\rL_7}&=\C7{\s^1\otimes[(-i\s^2)\otimes\s^3]},
  \quad&\quad
 \C8{\rL_8}&=\C8{\s^1\otimes[(-i\s^2)\otimes\s^1]}.
\end{alignat}
\end{subequations}
Noting the overall block-matrix structure of these matrices and that $\{\rL_2,\rL_2,\rL_5\}$ and $\{\rL_6,\rL_7,\rL_8\}$ generate two separate $\mathfrak{su}(2)$ algebras, a correspondence between $\rL_1,\{\rL_2,\rL_3,\rL_5\},\{\rL_6,\rL_7,\rL_8\},\rL_4$, respectively, to $\rD_+,\rD_+^{\sss\ha\hb},\rD_-^{\sss\ha\hb},\rD_-$ in \Eqs{MU1}{MU2} is strongly suggested.

Similar transformation relations and their explicit matrix representations can also be obtained from\eq{UM2}, using the well-known relationship between the super-differential operators $\rD_\rI$ and the supercharges $Q_\rI$\eq{eSSQD}:
\begin{equation}
   Q_\rI = i\rD_\rI + 2\d_{IJ}\,\q^\rJ\,\ddt,\qquad
   \rD_\rI =-iQ_\rI + 2i\d_{IJ}\,\q^\rJ\,\ddt,
 \label{eQDs}
\end{equation}
we easily obtain, for example:
\begin{alignat}{5}
  \rD_{\sss{\hK\,\ha}}{\bf A}
  &= i\,(\s^3)_{\sss\hK}{}^{\sss\hL}\,\bs{\J}_{\sss\hL\,\ha}
 \quad&&\To&\quad
  {\bf A}\big[\,-iQ_{\sss{\hK\,\ha}}
            +2i\,\d_{\sss\hK\hL}\d_{\sss\ha\hb}
                \q^{\sss\hL\,\hb}\,\big]
  &= i\,(\s^3)_{\sss\hK}{}^{\sss\hL}\,\bs{\J}_{\sss\hL\,\ha},
 \label{eD2Q.}\\
  &&&\To&\quad
  +\big[\,-iQ_{\sss{\hK\,\ha}}
         +2i\,\d_{\sss\hK\hL}\d_{\sss\ha\hb}
               \q^{\sss\hL\,\hb}\,\big]{\bf A}|
  &= i\,(\s^3)_{\sss\hK}{}^{\sss\hL}\,\bs{\J}_{\sss\hL\,\ha}|,
 \label{eD2Q}\\
 &&&\To&
  Q_{\sss{\hK\,\ha}}A
  &= -(\s^3)_{\sss\hK}{}^{\sss\hL}\,\j_{\sss\hL\,\ha},
 \label{eD2QF}
\end{alignat}
where ``$|$'' denotes setting $\q^\rI\to0$, which is how the additional, $\q$-dependent terms in\eq{eQDs} and in(\ref{eD2Q}) vanish by\eq{eD2QF}. The ``flip'' $\rD{\bf A}={\bf A}[-iQ{+}{\dots}]$ in\eq{eD2Q.} reflects the fact the $\rD$'s span left vector fields whereas the $Q$'s span right vector fields in superspace\cite{r1001}.
 In this manner, the system\eq{UM2} produces:
\begin{subequations}
 \label{UM2c}
\begin{alignat}{3}
 Q_{\sss{\hK\,\ha}}A
  &= -(\s^3)_{\sss{\hK}}{}^{\sss{\hL}}\,\j_{\sss{\hL\,\ha}},
 \quad&\quad
 Q_{\sss{\hK\hg}}A_{\sss{\ha\hb}}
  &= -\big[ \d_{\sss{\hg[\ha}}\,\j_{\sss{\hK\,\hb]}}
             +\tw\ell\,\ve_{\sss{\ha\hb\hg}}{}^{\sss\hd}\,
                        \j_{\sss{\hK\,\hd}}\,\big],\\[2mm]
 Q_{\sss{\hK\ha}}B
  &= -(\s^1)_{\sss{\hK}}{}^{\sss{\hL}}\,\j_{\sss{\hL\,\ha}},
 \quad&\quad
 Q_{\sss{\hK\hg}}B_{\sss{\ha\hb}}
  &= - \, \ve_{\sss{\hK\hL}}\, \big[ \d_{\sss{\hg[\ha}}\,\j_{\sss{\hL\,\hb]}}
             -\tw\ell\,\ve_{\sss{\ha\hb\hg}}{}^{\sss\hd}\,
                        \j_{\sss{\hL\,\hd}}\,\big],\\[2mm]
 Q_{\sss{\hK\ha}}\j_{\sss{\hL\,\hg}}
  &=\hbox to0mm{$\ddd- i\,\big[
      \d_{\sss{\ha\hg}}\,(\s^3)_{\sss{\hK\hL}}\,(\ddt A)
     +\d_{\sss{\hK\hL}}\,(\ddt A_{\sss{\ha\hg}})
     +\d_{\sss{\ha\hg}}\,(\s^1)_{\sss{\hK\hL}}\,(\ddt B)
     +\ve_{\sss{\hK\hL}}\,(\ddt B_{\sss{\ha\hg}})\big].$\hss}
 \quad&\quad
 &\mkern350mu
\end{alignat}
\end{subequations}

Owing to the connectivity of the super-differential constraints\eq{UM2}, the {\em\/a priori\/} unconstrained superfields in $({\bf A},{\bf B},{\bf A}_{\sss{\ha\hb}},{\bf B}_{\sss{\ha\hb}}|\bs{\J}_{\sss{\hK\,\ha}})$ contain no other component field than what appears in the final transformations such as\eq{UM2c}. To wit,
\begin{enumerate}\itemsep=-1pt\vspace{-3mm}
 \item Equations\eq{UM2} provide a supersymmetric mapping between 
 $\big\{{\bf A,B,A_{\sss\ha\hb},B_{\sss\ha\hb}}\big\}$ and
 $\big\{\bs{\J}_{\sss\hK\,\ha}\big\}$. This map is of maximum rank and is supersymmetric since the $\rD$'s anticommute with the $Q$'s, and so provides a supersymmetric isomorphism,
 $\Span({\bf A,B,A_{\sss\ha\hb},B_{\sss\ha\hb}}) \approx
   \Span(\bs{\J}_{\sss\hK\,\ha})$.
 \item Since $\wedge^r\q$-component fields in a superfield $\bs{X}$ are defined by evaluating
 $(\rD_{\rI_1}\cdots\rD_{\rI_r}\,\bs{X})$ at $\q=0$,
equations\eq{UM2.1} and\eq{UM2.2} imply that the $\wedge^{r+1}\q$-component fields within ${\bf A},{\bf B},{\bf A}_{\sss{\ha\hb}},{\bf B}_{\sss{\ha\hb}}$ are all $\wedge^r\q$-component fields within $\bs{\J}_{\sss{\hK\,\ha}}$, for $r\geq0$.
 \item In turn, equations\eq{UM2.3} imply that the $\wedge^{r+1}\q$-component fields within $\bs{\J}_{\sss{\hK\,\ha}}$ are all $\t$-derivatives of the $\wedge^r\q$-component fields within ${\bf A},{\bf B},{\bf A}_{\sss{\ha\hb}},{\bf B}_{\sss{\ha\hb}}$, for $r\geq0$.
 \item It follows that all component fields within the superfield multiplet $({\bf A},{\bf B},{\bf A}_{\sss{\ha\hb}},{\bf B}_{\sss{\ha\hb}}|\bs{\J}_{\sss{\hK\,\ha}})$ are either the lowest components listed in equations\eq{UM1.5} or their $\t$-derivatives.\QED
\end{enumerate}\vspace{-3mm}
The analogous applies to the super-differential system\eqs{MU1}{MU2}.

This proof in fact applies to all Superfield Adinkras: wherein every node in a given Adinkra is assigned an {\em\/a priori\/} unconstrained and unprojected superfield, and wherein every edge defines a pair of first order super-differential relations akin to the pairs exhibited in Eqs.\eq{UM2} or\eq{MU2}. This then provides a superfield representation of every one of the trillions of Adinkras and adinkraic supermultiplets classified by\cite{r6-3,r6-3.2}, and is in addition to the combined construction of Refs.\cite{r6-1,r6-1.2}.

\section {Supermultiplet Automorphisms}
 \label{a:Same}
Several distinct notions of {\em\/equivalence\/} between off-shell supermultiplets in general have been used in the literature, and it behooves to delineate the isomorphisms underlying such equivalences in our present study|and to highlight the differences between them.

In general, every supermultiplet is a {\em\/triple\/}, $(Q:{\bf R}_\f|{\bf R}_\j)$, where ${\bf R}_\f$ and ${\bf R}_\j$ are the vector spaces generated respectively by bosonic and fermionic component fields over a given spacetime and their spacetime derivatives, and a graded action of the supercharges, $Q$, generating the supersymmetry transformations between ${\bf R}_\f$ and ${\bf R}_\j$ and their derivatives. A general supersymmetric model will involve several supermultiplets, each of which having its own pair $({\bf R}_\f|{\bf R}_\j)$ equipped with a separate $Q$-action on this pair|but with the $Q$ operators themselves being common to all supermultiplets.

A field redefinition is a transformation of the pair $({\bf R}_\f|{\bf R}_\j)$, and is an {\em\/inner\/} transformation of each supermultiplet separately: the component fields of any supermultiplet can be redefined without any change induced in any other supermultiplet. By contrast, a transformation of the collection $\{Q_1,\cdots,Q_N\}$, such as a permutation or a linear combination, is common to all supermultiplets considered in a given model and so is an {\em\/outer\/} transformation for all of the involved supermultiplets: they are all affected by such a transformation.

For a transformation|inner or outer|of a supermultiplet to be an automorphism, its result must be {\em\/indistinguishable\/} from the original. It is this, inherently contextual nature of ``(in)dis\-tin\-guish\-a\-bility,'' that induces the various notions of (in)equivalence between two supermultiplets, and so renders all classification attempts just as context-sensitive.

The comparison of chiral and twisted-chiral supermultiplets in $(2,2)$-supersymmetry in $1{+}1$-dimensional spacetime (worldsheet) provide a well-known example\cite{rGHR}: In a certain basis of super-differential operators, these two complex superfields, $\F$ and $\X$ respectively, may be defined as satisfying the super-differential constraints
\begin{equation}
  [D_1-iD_2]\,\F=0=[D_3-iD_4]\,\F,\qquad\textit{vs.}\qquad
  [D_1-iD_2]\,\X=0=[D_3+iD_4]\,\X.
\end{equation}
Clearly, the transformation $D_4\to-D_4$ (and correspondingly $Q_4\to-Q_4$) swaps $\F\iff\X$. This being a bijection, it provides an isomorphism that renders each chiral superfield $\F$ equivalent to a twisted-chiral $\X$, and {\em\/vice versa\/}. Therefore, any model and action functional constructed with only twisted-chiral superfields can equally well be constructed with only chiral superfields, and there can be no physically observable distinction between them. However, $\F\iff\X$ is an outer automorphism and Ref.\cite{rGHR} shows that it is possible to construct a model and an action functional that indecomposably mixes $\F$ with $\X$, giving rise to target space geometries and field theory dynamics not achievable using only one or only the other type of superfields. The most celebrated consequence of the $\F\iff\X$ isomorphism between certain models with an unequal number of $\F$'s and $\X$'s is mirror symmetry\cite{rGP1,rMP1,rTwSJG1,rGLP,rMMYau1}.
 We thus refer to the chiral and twisted chiral superfields as equivalent (isomorphic), but {\em\/usefully distinct\/}.

Suppose now that the permissible action functionals were restricted to being at most bilinear in the supermultiplets. It is not hard to show that there exists no bilinear action functional mixing the chiral and twisted-chiral superfields indecomposably. This removes the means to physically distinguish one from the other supermultiplet, physical distinction being predicated on the observables in the theory and their dynamics|all of which depend on the choice of the action functional. Thus, whatever reasons that might have forced us to restrict action functionals to be bilinear would also render chiral and twisted chiral superfields indistinguishable|though they are not so in general.

In a qualitatively similar but technically subtler sense, there exist 30 distinct adinkraic $(8|8)$-dimensional supermultiplets of $N=8$ worldline supersymmetry, with the $E_8$ chromotopology\cite{r6-3}: Of the $8!$ column-permutations in the four binary 8-vectors at the right-hand side of Eqs.\eq{PE8}, many simply induce a permutation in the complete collection of binary linear combinations of the generators. As this complete collection is in fact the code, such a column-permutation turns out to be a symmetry of the code. The dimension of the permutation symmetry of the $e_8$ code being 1344, we are left with $\frac{8!}{1344}=30$ permutation-equivalent but distinct $e_8$ codes, and so also 30 equivalent but distinct projection systems\eq{PE8}, Adinkras\eq{eAB88}, and $(8|8)$-dimensional supermultiplets of $N=8$-extended worldline supersymmetry.

As discussed above, half of these are equivalent to each other by a linear field redefinition and we call these the ``ultra-multiplets,'' as are the other half which we call the ``twisted ultra-multiplets.'' 
 Indeed, $8\times8$ permutation matrices are in fact orthogonal, but half of them have determinant $-1$, the other half $+1$. While the second half therefore do form a subgroup of $\Spin(8)$, the full set ($8!=40,320$) of $8\times8$ permutation matrices|and therefore also those 30 that transform one $e_8$ code into another|form a discrete subgroup of the $\Pin(8)$ extension of $\Spin(8)$.

In turn, $\Spin(8)$ is unique in also possessing the well-known {\em\/triality\/}, which cycles ${\bf8}_v,{\bf8}_s$ and ${\bf8}_c$. Together with the $\Pin(8)/\Spin(8)=\ZZ_2$ reflections, this forms the $S_3$ {\em\/outer\/} automorphism of $\Spin(8)$, and it is possible to define the ``ultra-multiplet'' to include this full $S_3$-extension of $\Spin(8)$, denoted $\Spin(8)\ttmes S_3$, as its maximal group of symmetries. As pointed out, the particular assignment\eq{e888} was made up to this triality; any other assignment would serve just as well.

The various equivalences between distinct (twisted) ultra-multiplets may thus be layered:
\begin{enumerate}\vspace{-3mm}\itemsep=-1pt\setcounter{enumi}{-1}
 \item All $(8|8)$-dimensional supermultiplets of $N=8$ supersymmetry (\,=\,``ultra-multiplets'') are regarded equivalent:
 This level of (in)distinction corresponds to the situation where we consider a model constructed from perhaps several copies of the identically same Valise ultra-multiplet. In this case, a particular choice of a basis together with some of the choices made in specifying \Eqs{UM1}{UM2} and/or the assignments\eq{e888} will have been employed, but there would be no physically observable consequence of changing any and all of those by means of a $\Spin(8)\ttmes S_3$-transformation; no two distinct descriptions could result in any difference in the action functional|and so also amongst the observables of the system, so that all distinct descriptions would have to be regarded as physically equivalent.
 
 \item Ultra-multiplets are equivalent up to the triality of $\Spin(8)$, cycling the assignment of ${\bf R}_Q$, ${\bf R}_\f$ and ${\bf R}_\j$ to its three distinct irreducible 8-dimensional representations. Now, it is logically impossible to construct models with any two such distinct types of ultra-multiplets, since the supersymmetry of the model must be generated by $Q$'s that are common to all supermultiplets. Thus, ${\bf R}_Q$ must be common to the $Q$-action within all multiplets, which effectively rules out the possibility of mixing ultra-multiplets that differ by the triality of the assignment\eq{e888}. However, the triality of $\Spin(8)$ {\em\/is\/} involved in identifying $\Gf$ of non-Valise ultra-multiplets:
\begin{enumerate}\vspace{-3mm}\itemsep=-1pt
 \item[{\it I:}] with ${\bf R}_\f$ decomposing in the manner of the sequence\eq{eSeq}, as considered herein,
 \item[{\it II:}] with ${\bf R}_\j$ decomposing in the manner of the sequence\eq{eSeq}, as would be appropriate for the ``dark side'' of the ultra-multiplet family, including the incremental (fermionic) node-raises of the Adinkra\eq{eAF88},
 \item[{\it III:}] with ${\bf R}_Q$ decomposing in the manner of the sequence\eq{eSeq}, as in\eqs{MU1}{MU2}.
\end{enumerate}
While it is logically possible to mix the ultra-multiplets of type-I and type-II, those of type-III, as in \Eqs{MU1}{MU2}, cannot be mixed with either type-I or type-II, but provide a separate, alternate description|equipped with a simple sign-representation of ``twisting'' in \Eq{UTU}

 \item Ultra-multiplets, now with a fixed ${\bf R}_Q={\bf8}_v$ assignment, are equivalent up to the reflections in $\Pin(8)/\Spin(8)=\ZZ_2$. This swaps ${\bf R}_\f$ and ${\bf R}_\j$ as its two distinct 8-dimensional spinor irreps, one spanned by vectors in the root lattice that have an even number of negative coordinates, the other with an odd number. In the type-III description\eqs{MU1}{MU2}, this binary choice is represented the sign-choice in Eqs.\eq{UTU}. From the lessons of Ref.\cite{rGHR}, we expect that an action functional  mixing indecomposably the ultra-multiplet and its twisted variant would have to be a nonlinear (and/or gauged) $\s$-model, the target space of which would then admit a geometry not possible in any nonlinear and/or gauged $\s$-model constructed from only one of the two types.
 
 While a demonstration of the existence and explicit construction of such an action functional remains a tantalizing open problem, it is amusing to note that a bilinear mixing term cannot be $\Gf=\Spin(8)$-invariant: To see this, note that ${\bf R}_\f={\bf8}_s$ in the ultra-multiplet, while ${\bf R}_\vf={\bf8}_c$ in the twisted ultra-multiplet, while the fermions have the opposite assignments. Since ${\bf8}_s\otimes{\bf8}_c\not\supset\Ione$ in $\Spin(8)$, there can be no $\Spin(8)$-invariant bilinear mixing term suitable for a Lagrangian.

 \item Thirty ($30=\frac{8!}{1344}$) distinct ultra-multiplets are distinguished by distinct though $Q$-permutation equivalent $E_8$ topology. This implies the existence of a hierarchy of (presumably nonlinear and/or gauged) $\s$-models, indecomposably mixing between 2 and 30 of these variant (twisted) ultra-multiplets, and thus a class of target-space geometries not otherwise constructible. Thus, if no such action functionals can be constructed, the equivalences generated by Construction~4.2 of Ref.\cite{r6-3.2} are indeed physical equivalences; otherwise, these 30 incarnations of the (twisted) ultra-multiplet are all {\em\/usefully distinct\/}.
\end{enumerate}
\paragraph{Remark:}
The logical possibility of such a ``{\em\/useful distinction\/}'' implies that the total number of distinct supermultiplets increases from $\sim10^{12}$ without such distinction, to somewhere $\sim10^{47}$ with such distinction, and the number of usefully distinct models (``most general'' types of action functionals given a selection of supermultiplets) of $N\leq32$-extended supersymmetry to a combinatorially staggering number, somewhere ``log-log-halfway'' between Googol and Googolplex.
 Of course, most considered models also admit continuous parameters.

\section{Bases for Other Values of $N$}
 \label{a:Spin}
The effective symmetry of adinkraic supermultiplets essentially depends on three factors:
 ({\small\bf1})~the number of supercharges, $N$, acting within the supermultiplet,
 ({\small\bf2})~the degree, $k$, of the $(\ZZ_2)^k$-quotient chromotopology of the supermultiplet, and 
 ({\small\bf3})~the hanging, \ie, engineering unit assignments of its component fields. We focus herein on the {\em\/minimal\/} supermultiplets for given even $N$. These are maximally $(\ZZ_2)^k$-projected, with $k$ given by \Eq{eKmax}, which simplifies for even $N$:
\begin{equation}
   \vk(N)
   =\left\{\begin{array}{lcl}
             \fracm{N}2,    &\text{for} &N=0\text{ (mod}~8),\\[2mm]
             \fracm{N}2-1,&\text{for} &N=2,4,6\text{ (mod}~8),
           \end{array}\right.
 \label{eKmax2}
\end{equation}
resulting in
\begin{align}
 \dim(\textit{min.~supermultiplet\/})
  &=\left\{\begin{array}{lcl}
              (2^{\frac{N}2-1}|2^{\frac{N}2-1}),
                  &\text{for} &N=0\text{ (mod}~8),\\[2mm]
              (2^{\frac{N}2}|2^{\frac{N}2}),
                  &\text{for} &N=2,4,6\text{ (mod}~8),
           \end{array}\right.
\intertext{with the two powers of 2 denoting the numbers of bosonic and fermionic degrees of freedom, respectively. The dimension of the minimal spinor representations of $\Spin(2n)$ being $2^{n-1}$, this suggests:}
 \Gf(\textit{min.~supermultiplet})
  &\subseteq\left\{\begin{array}{lcl}
              \Spin(N),    &\text{for} &N=0\text{ (mod}~8),\\[2mm]
              \Spin(N{+}2),&\text{for} &N=2,4,6\text{ (mod}~8).
           \end{array}\right.
 \label{eGeff}
\end{align}
With these assignments, note that the defining $n$-dimensional vector representation of $\Spin(n)$ must provide the mapping $\bs{n}_v:(\bs{2^{n-1}})_s\iff(\bs{2^{n-1}})_c$ and in agreement with the conditions\eq{eReff}. For $N=0\pmod8$, $\Span(Q_1,\cdots,Q_N)$ precisely suffices, but in the $N=2,4,6\pmod8$ cases, either the $(N{+}2)$-component vector representation would have to be spanned by a suitable subset of $\{Q_1,\cdots,Q_N,\rD_1,\cdots,\rD_N\}$, or $\Gf\subseteq\Spin(N)\subset\Spin(N{+}2)$, restricting from\eq{eGeff}.

While this suggests generalizations of the present results to the $N=2,4,6\pmod8$ cases, we defer their discussion to a future opportunity, focus on $N=0\pmod8$ cases and fix $n:=N/2$.

\paragraph{The $\textsl{\bfseries U}(1)\to\mathbb{Z}_2^e$ Charges}
The $\sU(1)$ charge-assignment for the subgroup $\SU(n){\times}\sU(1)\subset\Spin(2n)$ in equation 
(\ref{eSpin(2n)}) and\eq{eSpin(2n+1)} follows the pattern:
\begin{equation}
 \vC{\begin{picture}(140,20)(-40,-2)
      \put(-40,0){$(\bs{2^{2n-1}})_\text{co-spinor}$}
       \put(-12,2){\vector(1,0){8}}
       \put(-33,5){\vector(0,1){8}}
       \put(-33,5){\vector(0,-1){0}}
        \put(-32,8){\footnotesize$(\bs{2n})_\text{vector}$}
      \put(-40,15){$(\bs{2^{2n-1}})_\text{spinor}$}
       \put(-12,17){\vector(1,0){8}}
      \put(0,0){$\bs{\binom{n}{0}}_{q_0}$}
       \put(6,5){\vector(3,2){10}}
        \put(5,9){\footnotesize$\bs{n}_{+1}$}
       \put(17,11){\vector(-3,-2){10}}
        \put(13,5){\footnotesize$\bs{n^*}\!\!_{-1}$}
      \put(19,15){$\bs{\binom{n}{1}}_{q_1}$}
       \put(39,5){\vector(-3,2){10}}
        \put(35,9){\footnotesize$\bs{n^*}\!\!_{-1}$}
       \put(28,11){\vector(3,-2){10}}
        \put(27,5){\footnotesize$\bs{n}_{+1}$}
      \put(40,0){$\bs{\binom{n}{2}}_{q_2}$}
       \put(46,5){\vector(3,2){10}}
        \put(45,9){\footnotesize$\bs{n}_{+1}$}
       \put(57,11){\vector(-3,-2){10}}
        \put(53,5){\footnotesize$\bs{n^*}\!\!_{-1}$}
      \put(59,15){$\bs{\binom{n}{3}}_{q_3}$}
       \put(79,5){\vector(-3,2){10}}
        \put(75,9){\footnotesize$\bs{n^*}\!\!_{-1}$}
       \put(68,11){\vector(3,-2){10}}
        \put(67,5){\footnotesize$\bs{n}_{+1}$}
      \put(80,0){$\bs{\binom{n}{4}}_{q_4}$}
      \put(87,8){\Large$\cdots$}
     \end{picture}}
 \label{eZZ}
\end{equation}
whereby
\begin{equation}
  q_p = q_0+p.
\end{equation}
Since $\sU(1)\subset\Spin(2n)$ and the whole array is a complete representation of $\Spin(2n)$, the trace of the $\sU(1)$ generator must vanish, \ie, the charges must add up to zero:
\begin{equation}
  0=\sum_{p=0}^n q_p=(n{+}1)q_0+\binom{n+1}2,\quad\To\quad q_0=-\inv2n,
\end{equation}
and the $q_p$ range $\{-\frc{n}2,-\frc{n-2}2,\cdots,+\frc{n_2}2,+\frc{n}2\}$|exactly as do the (angular momentum) $\hat{J}_3$ eigenvalues in quantum mechanics!
 In fact, the charges add up to zero both separately in the top and the bottom row of\eq{eZZ}, owing to the identities
\begin{equation}
 \sum_{p\text{ even}}(p-\inv2n)\binom{n}{p}=0,\quad\text{and}\quad
 \sum_{p\text{ odd}}(p-\inv2n)\binom{n}{p}=0,
\end{equation}
in accord with the fact that the top row of\eq{eZZ} by itself displays the $\SU(n){\times}\sU(1)\subset\Spin(2n)$ decomposition of the ${\bf R}_s$ representation of $\Spin(2n)$, and the bottom row displays the corresponding decomposition of ${\bf R}_c$, wherein the trace of $\sU(1)\subset\Spin(2n)$ must be zero.

In turn, we can represent the spinors and the co-spinors of $\Spin(2n)$ in its root lattice as spanned by $n$-vectors $(\pm\inv2,\cdots,\pm\inv2)$ with, respectively an even and an odd number of negative components. These are easily partitioned into $\binom{p}2$ such vectors with $p$ negative signs, and $q_p$ in\eq{eZZ} are then simply the sums of components for each such partition.

When passing to the real subgroup $\Spin(n){\times}\ZZ_2=\Ree\big(\SU(n){\times}\sU(1)\big)$, the complex distinction between $\bs{n}$ and $\bs{n^*}$ is lost, they become interchangeable in the diagram
(\ref{eZZ}). Thus, for $n=0\pmod4$, we may shift (the remnant of) the $\sU(1)$ charge so that:
\begin{equation}
 \vC{\begin{picture}(160,20)(0,-2)
      \put(0,0){$\bs{\binom{n}{0}}_{0}$}
       \put(6,5){\vector(3,2){10}}
        \put(4,9){\footnotesize$\bs{n}_{+1}$}
       \put(17,11){\vector(-3,-2){10}}
        \put(13,5){\footnotesize$\bs{n}_{-1}$}
      \put(20,15){$\bs{\binom{n}{1}}_{+1}$}
       \put(39,5){\vector(-3,2){10}}
        \put(35,9){\footnotesize$\bs{n}_{+1}$}
       \put(28,11){\vector(3,-2){10}}
        \put(27,5){\footnotesize$\bs{n}_{-1}$}
      \put(40,0){$\bs{\binom{n}{2}}_0$}
       \put(46,5){\vector(3,2){10}}
        \put(44,9){\footnotesize$\bs{n}_{+1}$}
       \put(57,11){\vector(-3,-2){10}}
        \put(53,5){\footnotesize$\bs{n}_{-1}$}
      \put(60,15){$\bs{\binom{n}{3}}_{+1}$}
      \put(67,8){\Large$\cdots$}
      \put(74,0){$\bs{\binom{n}{n/2}}_0$}
      \put(83,8){\Large$\cdots$}
      \put(90,15){$\bs{\binom{n}{n-3}}_{-1}$}
       \put(111,5){\vector(-3,2){10}}
        \put(107,9){\footnotesize$\bs{n}_{-1}$}
       \put(100,11){\vector(3,-2){10}}
        \put(99,5){\footnotesize$\bs{n}_{+1}$}
      \put(110,0){$\bs{\binom{n}{n-2}}_0$}
       \put(120,5){\vector(3,2){10}}
        \put(118,9){\footnotesize$\bs{n}_{-1}$}
       \put(131,11){\vector(-3,-2){10}}
        \put(127,5){\footnotesize$\bs{n}_{+1}$}
      \put(131,15){$\bs{\binom{n}{n-1}}_{-1}$}
       \put(155,5){\vector(-3,2){10}}
        \put(151,9){\footnotesize$\bs{n}_{-1}$}
       \put(144,11){\vector(3,-2){10}}
        \put(143,5){\footnotesize$\bs{n}_{+1}$}
      \put(156,0){$\bs{\binom{n}{n}}_0$}
     \end{picture}}
 \label{eZZ0}
\end{equation}
The $n{\times}n$ metric of the real subgroup $\Spin(n){\times}\ZZ_2=\Ree\big(\SU(n){\times}\sU(1)\big)$ induces the decomposition of $\binom{n}{n/2}_0$ into its self-dual and anti-self-dual parts, and the entire sequence\eq{eZZ0} exhibits a left-right reflection symmetry across the middle. The so-shifted $U(1)$ charge is the assignment used in equation\eq{BQ} and\eq{Bj}, which agrees with the labeling carried by the $\hK$-type indices and corresponds to what we denoted above as $\ZZ_2^e$, and also with the so-denoted ``charges'' appearing in Table~\ref{t:O8} and Figure~\ref{f:SO8}. This shift can also be traced through the exponential mapping\eq{exp}.

\paragraph{Particular Cases{\rm:}}
For the $N=8$ ultra-multiplet, Table~\ref{t:O8} shows that the basis of Ref.\cite{rGR0} corresponds to the decomposition
\begin{equation}
\begin{aligned}
  {\bf R}_Q &=({\bf2,2})_-\oplus({\bf2,2})_+\\
  {\bf R}_\f&=({\bf1,1})_0\oplus({\bf3,1})_0\oplus({\bf1,3})_0\oplus({\bf1,1})_0\\
  {\bf R}_\j&=({\bf2,2})_+\oplus({\bf2,2})_-
\end{aligned}
 ~\Bigg\}\quad\text{of }\SZ^e.
 \label{eN8}
\end{equation}
For $N\geq8$, such decompositions may be less useful, since the zig-zag diagram\eq{eZZ0} will not be able to accommodate the raising of an arbitrary number of nodes. As an example, consider the minimal, $(128|128)$-dimensional supermultiplet of $N=16$ extended worldline supermultiplet:
\begin{equation}
\begin{aligned}
  {\bf R}_Q &={\bf8}_-\oplus{\bf8}_+\\
  {\bf R}_\f&={\bf1}_0\oplus{\bf28}_0\oplus{\bf35}^+_0\oplus{\bf35}^-_0\oplus{\bf28}_0\oplus{\bf1}_0\\
  {\bf R}_\j&={\bf8}_+\oplus{\bf56}_+\oplus{\bf56}_-\oplus{\bf8}_-
\end{aligned}
 ~\Bigg\}\quad\text{of }\Spin(8){\times}\ZZ_2^e.
 \label{eN16}
\end{equation}
This may be obtained from the unprojected, $(2^{15}|2^{15})=(32{,}768\,|\,32{,}768)$-dimensional $N=16$ supermultiplet, {\em\/via\/} eight successive, commuting $\ZZ_2$-projections and permits the description of raising $1,2,28,29,30,\cdots$, but not $3,4,\cdots,27$ nodes, \etc,|without changing to a basis wherein the $\Spin(8){\times}\ZZ_2^e$ is manifestly broken to a subgroup.

We pause to note that the numbers in the $N=16$ decomposition\eq{eN16} will seem extremely familiar to aficionados of supergravity theory, reminding of the multiplicities of bosonic (top row) and fermionic (bottom row) helicity states in 4D, ${\cal N}=8$ supergravity.

However, with a total of $N=32$ supercharges, the worldline shadow of the {\em\/a priori\/} unconstrained and unprojected supermultiplet of 4D $N=8$ has $2^{31}+2^{31}$ components, and admits $2{\times}85=170$ distinct\Ft{These supermultiplets have 85 distinct chromotopology types, each with two inequivalent choices of edge-dashing\cite{r6-3,r6-3.2}.} $(\ZZ_2)^{16}$-projections\cite{r6-3,r6-3.2}. This leaves 170 distinct $2^{15}+2^{15}$-component supermultiplet classes|each of which admitting a combinatorially enormous number of distinct ``hangings,'' \ie, distinct possible assignments of engineering units to the component fields.
 The simplest of these, the Valise hangings of such supermultiplets form only two equivalence classes that may be parametrized in a fashion straightforwardly generalizing the ultra-multiplet\eqs{UM1}{UM2}, complete with $\tl=\pm1$ labeling the two equivalence classes. This then affords an organization of this combinatorially enormous family of supermultiplets in the ``root superfield'' manner\cite{rGLP}, as proven in Ref.\cite{r6-1}.
 This Valise hanging of the $(2^{15}|2^{15})=(32{,}768\,|\,32{,}768)$-component worldline $N=32$-supersymmetric supermultiplet has the component fields form
\begin{equation}
\begin{aligned}
  {\bf R}_Q &={\bf16}_-\oplus{\bf16}_+\\
  {\bf R}_\f&={\bf1}_0\oplus{\bf120}_0\oplus
              {\bf1820}_0\oplus{\bf8008}_0\oplus{\bf6435}^+_0\oplus
              {\bf6435}^-_0\oplus{\bf8008}_0\oplus{\bf1820}_0\oplus
              {\bf120}_0\oplus{\bf1}_0\\
  {\bf R}_\j&={\bf16}_+\oplus{\bf560}_+\oplus{\bf4368}_+\oplus{\bf11440}_+\oplus
              {\bf11440}_-\oplus{\bf4368}_-\oplus{\bf560}_-\oplus{\bf16}_-
\end{aligned}
 \label{eN32}
\end{equation}
of $\Spin(16){\times}\ZZ_2^e$, in a straightforward generalization of the ultra-multiplet parametrization in Ref.\cite{rGR0}.

Indeed, the numbers in the decomposition\eq{eN32} look nothing like the multiplicities in the familiar 4-dimensional, ${\cal N}=8$ supergravity. This is simply because the decompositions\eqs{eN8}{eN32} are based on the symmetry group that is, for each $N$, common to a large class of different ``hangings'' of the described {\em\/worldline\/} supermultiplets, many of which most probably do not have a dimensional oxidization to 4-dimensional spacetime supermultiplets. Therefore, the $\Spin(n/2){\times}\ZZ_2^e$-basis described herein and implicit in the decompositions\eqs{eN8}{eN32} is most certainly not ``aligned'' with an embedded 4-dimensional Lorentz group, $\Spin(1,3)$, in $\max(\Gf)$. Furthermore, unlike $\ddt$ on which $\Spin(1)=\ZZ_2$ acts trivially, $\Spin(1,3)$ acts far from trivially on the energy-momentum 4-vector in 4-dimensional spacetime; this forces a modification in the 4-dimensional spacetime analogues of Eqs.\eq{eReff}, so also in the assignments\eq{e888}, and thereby the very definition of $\Gf$ itself.

\section{The $E_8$ Algebra, Root Lattice Bases, Code and Adinkra Chromotopology}
 \label{a:E8}
This appendix collects a telegraphic review of the correspondences between: ({\small\bf1})~the $E_8$ algebra, ({\small\bf2})~root lattice bases, ({\small\bf3})~$e_8$ doubly-even binary linear block codes, and ({\small\bf4})~adinkraic supermultiplets with $E_8$ chromotopology\cite{r6-3,r6-3.2,r6-3c}. Information on the $E_8$ algebra and root lattices may be found in many texts on Lie algebras; see {\em\/e.g.\/}, Ref.\cite{rWyb,rJFA} and the on-line summary\cite{rWP-E8}. The correspondence between the $E_8$ root lattice bases and the $e_8$ binary code follows the so-called ``Construction~A''\cite{rCPS}.

\paragraph{The $E_8$ Algebra{\rm,}} familiar from its many diverse applications in mathematical physics, has a $\spin(16)$ maximal, regular subalgebra. With respect to this, the $E_8$ adjoint representation decomposes as $\bf248\to120\oplus128$, where ${\bf120}\sim\Span(J_{[ab]},~a,b=1,\cdots,16)$ and ${\bf128}\sim\Span(Q^{}_A=Q^\dagger_A,~A=1,\cdots,128)$, satisfying:
\begin{subequations}
 \label{eE8a}
\begin{gather}
 \big[\,J_{[ab]}\,,\,J_{[cd]}\,\big]
  =\d_{ad}J_{[bc]}-\d_{ac}J_{[bd]}-\d_{bd}J_{[ac]}+\d_{bc}J_{[ad]},\\
 \big[\,J_{[ab]}\,,\,Q_A\,\big]=\fracm12\,(\G_{[a|})_A{}^B\,(\G_{|b]})_B{}^C\,Q_C,\qquad
 \big[\,Q_A\,,\,Q_B\,\big]=(\G^{[a|})_A{}^C\,(\G^{|b]})_C{}^D\,\d_{DB}\,J_{[ab]},
\end{gather}
\end{subequations}
where $(\G_a)_A{}^B$ are 16 suitable $128{\times}128$ Dirac matrices, $\d_{ab}$ the defining metric of $\spin(16)$, and $\d_{AB}$ the metric on its 128-dimensional real spinor representation.

\paragraph{The $E_8$ Root Lattice{\rm,}} $\L_{E_8}$, of the $E_8$ algebra\eq{eE8a} consists of the eight mutually commuting Cartan elements $\{J_{[1\,2]},J_{[3\,4]},\cdots,J_{[15\,16]}\}$, represented as all coinciding with the point of origin in an 8-dimensional Euclidean space. The remaining 112 elements among $\{J_{[ab]},~a,b=1,\cdots,16\}$ then correspond to integral 8-tuples that are permutations of $(\pm1,\pm1,0,\cdots,0)$. The $Q_A$ in turn correspond to half-integral 8-tuples of the form $(\pm\inv2,\cdots,\pm\inv2)$|with either an even or an odd total number of positive/negative components|corresponding to a choice of the even/odd $E_8$ class of lattices, $\L_{E_8}^+$ {\em\/vs\/}.\ $\L_{E_8}^-$.
 The Euclidean length of all 240 root-vectors in both $\L_{E_8}^+$ {\em\/vs\/}.\ $\L_{E_8}^-$ is $\sqrt{2}$, and the lattices $\L_{E_8}^+,\L_{E_8}^-$ are in fact isomorphic.

A convenient basis for $\L_{E_8}^+$|one of many|consists of the 8-tuples listed here as the rows of the matrix:
\begin{equation}
  \left[\begin{array}{r}
   \bs{\a}_1\\
   \bs{\a}_2\\
   \bs{\a}_3\\
   \bs{\a}_4\\
   \bs{\a}_5\\
   \bs{\a}_6\\
   \bs{\a}_7\\
   \bs{\a}_8\\
        \end{array}\right]
  :=
  \left[\begin{array}{@{\>}r@{~}r@{~}r@{~}r@{~}r@{~}r@{~}r@{~}r}
    \inv2  &-\inv2 &-\inv2 &-\inv2 &-\inv2 &-\inv2 &-\inv2 &+\inv2\\
   -1      &1      &0      &0      &0      &0      &0      &0\\
    0      &-1     &1      &0      &0      &0      &0      &0\\
    0      &0      &-1     &1      &0      &0      &0      &0\\
    0      &0      &0      &-1     &1      &0      &0      &0\\
    0      &0      &0      &0      &-1     &1      &0      &0\\
    0      &0      &0      &0      &0      &-1     &1      &0\\
    1      &1      &0      &0      &0      &0      &0      &0\\
        \end{array}\right]
  \qquad\longleftrightarrow\qquad
  \vC{\unitlength=.95mm
      \begin{picture}(20,42)
       \put(0,42){\small$\bs{\a}_1$}
        \put(1.2,37.5){\line(0,1){3.5}}
       \put(0,35){\small$\bs{\a}_2$}
        \put(1.2,30.5){\line(0,1){3.5}}
       \put(0,28){\small$\bs{\a}_3$}
        \put(1.2,23.5){\line(0,1){3.5}}
       \put(0,21){\small$\bs{\a}_4$}
        \put(1.2,16.5){\line(0,1){3.5}}
       \put(0,14){\small$\bs{\a}_5$}
        \put(1.2,9.5){\line(0,1){3.5}}
       \put(0,7){\small$\bs{\a}_6$}
        \put(1.2,2.5){\line(0,1){3.5}}
       \put(0,0){\small$\bs{\a}_7$}
        \put(5,29){\line(1,0){3.5}}
       \put(9,28){\small$\bs{\a}_8$}
      \end{picture}}
 \label{eE8b}
\end{equation}
where a link between $\bs{\a}_i$ and $\bs{\a}_j$ in the Dynkin diagram to the right indicates $\bs{\a}_i\bs{\cdot}\bs{\a}_j=-1$, and the absence of a link indicates $\bs{\a}_i\bs{\cdot}\bs{\a}_j=0$. Then, $\L_{E_8}^+$ is the collection of integral multiples of $\{\bs{\a}_1,\cdots,\bs{\a}_8\}$, and $\L_{E_8}^-$ is obtained by changing the sign in the entries in any odd number of columns of the $8{\times}8$ matrix formed by the eight row-vectors\eq{eE8b}.

\paragraph{An Integral-Length Rescaling:} The $\L_{E_8}$ lattice may be effectively rescaled {\em\/via\/} left-multiplication of the $8{\times}8$ matrix formed by the eight row-vectors\eq{eE8b} by $\IS=(\bs{\s}_1{+}\bs{\s}_3){\times}\Ione_4$:
\begin{equation}
 \hbox{\scriptsize$\ddd
  \left[\begin{array}{r@{~}r@{~~~}r@{~}r@{~~~}r@{~}r@{~~~}r@{~}r}
         1 & 1 &0 &0  &0 &0  &0 &0 \\
         1 &-1 &0 &0  &0 &0  &0 &0 \\
         0 &0  &1 & 1 &0 &0  &0 &0 \\
         0 &0  &1 &-1 &0 &0  &0 &0 \\
         0 &0  &0 &0  &1 & 1 &0 &0 \\
         0 &0  &0 &0  &1 &-1 &0 &0 \\
         0 &0  &0 &0  &0 &0  &1 & 1\\
         0 &0  &0 &0  &0 &0  &1 &-1\\
        \end{array}\right]
  \left[\begin{array}{@{~}r@{~~}r@{~~}r@{~~}r@{~~}r@{~~}r@{~~}r@{~~~}r}
    \tfrac12 &-\tfrac12 &-\tfrac12 &-\tfrac12 &-\tfrac12 &-\tfrac12 &-\tfrac12 & \tfrac12\\
    -1 & 1 & 0 & 0 & 0 & 0 & 0 & 0\\
     0 &-1 & 1 & 0 & 0 & 0 & 0 & 0\\
     0 & 0 &-1 & 1 & 0 & 0 & 0 & 0\\
     0 & 0 & 0 &-1 & 1 & 0 & 0 & 0\\
     0 & 0 & 0 & 0 &-1 & 1 & 0 & 0\\
     0 & 0 & 0 & 0 & 0 &-1 & 1 & 0\\
     1 & 1 & 0 & 0 & 0 & 0 & 0 & 0\\
        \end{array}\right]
  =
  \left[\begin{array}{@{~}r@{~~}r@{~~}r@{~~}r@{~~}r@{~~}r@{~~~}r@{~~}r}
    0 & 1 &-1 & 0 &-1 & 0 & 0 &-1\\
    0 &-2 & 0 & 0 & 0 & 0 & 0 & 0\\
   -1 & 1 & 1 & 1 & 0 & 0 & 0 & 0\\
    0 & 0 & 0 &-2 & 0 & 0 & 0 & 0\\
    0 & 0 &-1 & 1 & 1 & 1 & 0 & 0\\
    0 & 0 & 0 & 0 & 0 &-2 & 0 & 0\\
    0 & 0 & 0 & 0 &-1 & 1 & 1 & 1\\
    2 & 0 & 0 & 0 & 0 & 0 & 0 & 0\\
        \end{array}\right]$}
 \label{eE8c}
\end{equation}
The row-vectors in the resulting matrix span a lattice isomorphic to $\L_{E_8}^+$ but now the minimal non-zero vectors have Euclidean length 2. Thus, the Cartan matrix,
 $A_{ij}:=2\frac{(\bs{\a}_i,\bs{\a}_j)}{(\bs{\a}_i,\bs{\a}_i)}$, computed from\eq{eE8c} equals the one computed from\eq{eE8a}. The binary, \ie, (mod~2) reduction of the matrix\eq{eE8c} becomes
\begin{equation}
  \hbox{\scriptsize$\ddd\begin{bmatrix}
   0 &1 &1 &0 &1 &0 &0 &1\\
   0 &0 &0 &0 &0 &0 &0 &0\\
   1 &1 &1 &1 &0 &0 &0 &0\\
   0 &0 &0 &0 &0 &0 &0 &0\\
   0 &0 &1 &1 &1 &1 &0 &0\\
   0 &0 &0 &0 &0 &0 &0 &0\\
   0 &0 &0 &0 &1 &1 &1 &1\\
   0 &0 &0 &0 &0 &0 &0 &0\\
  \end{bmatrix}$}
 \quad\longrightarrow\quad
  \begin{bmatrix}
   0 &1 &1 &0 &1 &0 &0 &1\\
   1 &1 &1 &1 &0 &0 &0 &0\\
   0 &0 &1 &1 &1 &1 &0 &0\\
   0 &0 &0 &0 &1 &1 &1 &1\\
  \end{bmatrix}
  =:
  \begin{bmatrix}
    \bs{b}_1\\
    \bs{b}_2\\
    \bs{b}_3\\
    \bs{b}_4\\
  \end{bmatrix}
 \label{eE8C}
\end{equation}
the rows of which generate the doubly-even binary linear block code $e_8$, which consists of all binary linear combinations $\vec{\bs{b}}:=\b^i\bs{b}_i$, where $\b^i\in\{0,1\}$ and bitwise summation is implicit. All 16 so-obtained binary 8-vectors have a doubly-even Hamming weight, \ie, the sum of 1's adds up to $0\pmod4$; they are all mutually orthogonal, $\vec{\bs{b}}\cdot\vec{\bs{b}}{}'=0\pmod2$ and self-orthogonal, $\vec{\bs{b}}\cdot\vec{\bs{b}}=0\pmod2$.

\paragraph{From Code to a Lattice Basis:}
Conversely, given a set of generators of the doubly-even code $e_8$, we reconstruct a $\L_{E_8}$-basis by judiciously toggling the sign of a few bits so the 8-vectors become mutually orthogonal without the (mod~2) reduction, and then add three 8-vectors: each with a single non-zero $\pm2$ entry\Ft{The (mod~2) reduction of each such vector is the binary null-vector, contained in every binary code.}, so that the Euclidean scalar product of each row-vector with both its predecessor and its successor is $-2$, and zero otherwise. For example:
\begin{equation}
 \hbox{\scriptsize$\ddd
  \left[\begin{array}{r@{~}r@{~}r@{~}r@{~}r@{~}r@{~}r@{~}r}
    1 & 1 & 1 & 1 & 0 & 0 & 0 & 0\\
    0 & 0 & 1 & 1 & 1 & 1 & 0 & 0\\
    0 & 0 & 0 & 0 & 1 & 1 & 1 & 1\\
    1 & 0 & 1 & 0 & 1 & 0 & 1 & 0\\
        \end{array}\right]$}
 ~\longrightarrow~
 \hbox{\scriptsize$\ddd
  \left[\begin{array}{@{\,}r@{~}r@{~}r@{~}r@{~}r@{~}r@{~}r@{~}r}
    1 & 1 & 1 &-1 & 0 & 0 & 0 & 0\\
    0 & 0 &-1 &-1 &-1 & 1 & 0 & 0\\
    0 & 0 & 0 & 0 & 1 & 1 & 1 &-1\\
   -1 & 0 & 1 & 0 &-1 & 0 & 1 & 0\\
        \end{array}\right]$}
 ~\longrightarrow~
 \hbox{\scriptsize$\ddd
  \left[\begin{array}{@{\,}r@{~}r@{~}r@{~}r@{~}r@{~}r@{~}r@{~}r}
    1 & 1 & 1 &-1 & 0 & 0 & 0 & 0\\
    0 & 0 & 0 & 2 & 0 & 0 & 0 & 0\\
    0 & 0 & 1 &-1 & 1 & 1 & 0 & 0\\
    0 & 0 & 0 & 0 & 0 &-2 & 0 & 0\\
    0 & 0 & 0 & 0 & 1 & 1 & 1 &-1\\
    0 & 0 & 0 & 0 & 0 & 0 &-2 & 0\\
   -1 & 0 & 1 & 0 &-1 & 0 & 1 & 0\\
        \end{array}\right]$}.
 \label{eE8d}
\end{equation}
Finally, we add the eighth 8-vector with a single non-zero $\pm2$ entry so that its Euclidean scalar product with either the $3^\text{rd}$ or $5^\text{th}$ row-vector is $-2$, but vanishes with all other row-vectors:
\begin{equation}
 \hbox{\scriptsize$\ddd
  \left[\begin{array}{r@{~}r@{~}r@{~}r@{~}r@{~}r@{~}r@{~}r}
    1 & 1 & 1 & 1 & 0 & 0 & 0 & 0\\
    0 & 0 & 1 & 1 & 1 & 1 & 0 & 0\\
    0 & 0 & 0 & 0 & 1 & 1 & 1 & 1\\
    1 & 0 & 1 & 0 & 1 & 0 & 1 & 0\\
        \end{array}\right]$}
 ~\longrightarrow~
 \hbox{\scriptsize$\ddd
  \left[\begin{array}{@{\,}r@{~}r@{~}r@{~}r@{~}r@{~}r@{~}r@{~}r}
    1 & 1 & 1 &-1 & 0 & 0 & 0 & 0\\
    0 & 0 & 0 & 2 & 0 & 0 & 0 & 0\\
    0 & 0 &-1 &-1 &-1 & 1 & 0 & 0\\
    0 & 0 & 0 & 0 & 0 &-2 & 0 & 0\\
    0 & 0 & 0 & 0 & 1 & 1 & 1 &-1\\
    0 & 0 & 0 & 0 & 0 & 0 &-2 & 0\\
   -1 & 0 & 1 & 0 &-1 & 0 & 1 & 0\\
    0 & 0 & 0 & 0 & 0 & 0 & 0 & 2\\
        \end{array}\right]$}
 ~\tooo{\>\bs{\cdot}\,\IS/2\>}~
 \hbox{\scriptsize$\ddd
  \left[\begin{array}{@{\,}r@{~}r@{~}r@{~}r@{~}r@{~}r@{~}r@{~}r}
    1 & 0 & 0 & 1 & 0 & 0 & 0 & 0\\
    0 & 0 & 1 &-1 & 0 & 0 & 0 & 0\\
    0 & 0 &-1 & 0 & 0 &-1 & 0 & 0\\
    0 & 0 & 0 & 0 &-1 & 1 & 0 & 0\\
    0 & 0 & 0 & 0 & 1 & 0 & 0 & 1\\
    0 & 0 & 0 & 0 & 0 & 0 &-1 &-1\\
   -\tfrac12 &-\tfrac12 & \tfrac12 & \tfrac12 &-\tfrac12 &-\tfrac12 & \tfrac12 & \tfrac12\\
    0 & 0 & 0 & 0 & 0 & 0 & 1 &-1\\
        \end{array}\right]$}.
 \label{eE8f}
\end{equation}
After toggling the signs of the elements in the $2^\text{nd},3^\text{rd},4^\text{th}$ and $8^\text{th}$ column, this is a row- and column-permutation of the $\L_{E_8}^+$-basis\eq{eE8b}; by toggling back the sign of the element in the $2^\text{nd}$ column, say, one obtains a $\L_{E_8}^-$-basis. As noted above, all $\L_{E_8}^+$ and $\L_{E_8}^-$ lattices are isomorphic as lattices. The above construction of a $\L_{E_8}$-basis proceeds with a clear aim to recover a lattice basis to correspond to the Dynkin diagram\eq{eE8b}. It would be interesting to prove that the ultimate result, the generation of an $E_8$-lattice from the $e_8$ code, does not depend on the choices made herein; this is however beyond the scope of this case-study.

\paragraph{Ultra-multiplets:}
On the other hand, the exponential map\eq{exp} corresponds a quotient of the 8-cube by the binary code $e_8$ to the projection of the {\em\/a priori\/} unconstrained and unprojected $(128|128)$-dimensional Valise supermultiplet of $N{=}8$ worldline supersymmetry, with the chromotopology of the 8-cube, $I^8$, and to the ultra-multiplet\eqs{UM1}{UM2} with the $E_8=I^8/e_8$ chromotopology\cite{r6-3,r6-3.2}. That is, the exponential map\eq{exp} assigns to every binary 8-vector $\vec{\bs{b}}\in e_8$ a $Q$-monomial, $\bs{Q^{\vec{b}}}$, the action of which on every component field of the ultra-multiplet is equivalent to $H^{\frac12\wt(\vec{\bs{b}})}$.

Throughout\eqs{eE8b}{eE8f}, the $\rI^\text{th}$ column in each matrix corresponds to the $\rI^\text{th}$ supercharge, and toggling the sign of the matrix elements in\eqs{eE8b}{eE8c} within any such column may be corresponded to changing the sign of $Q_\rI$. In turn then, toggling the sign in any {\em\/odd\/} number of columns corresponds to toggling the sign of an odd number of supercharges, which in turn has the effect of toggling between the ultra-multiplet family\eqs{UM1}{UM2} and its twisted variant|the Klein-flip of \Eqs{UM1}{UM2}, or alternatively, toggling between the two sign options in \Eqs{MU1}{MU2}. The rows in the right-hand side matrix\eq{eE8c} itself thus correspond to $Q$-monomials the action of which is equivalent to $H^{\frac12\wt(\vec{\bs{b}})}$ upon each component field of the ultra-multiplet: the rows with single $\pm2$ entries satisfy this trivially, since $\pm(Q_\rI)^2=\pm H$ already|corresponding to the fact that such rows reduce, (mod~2), to null-vectors in the binary code $e_8$.

In this way, the $\L_{E_8}^\pm$-lattice basis matrices\eqs{eE8b}{eE8c} themselves are seen to encode the quotient chromotopology of the ultra-multiplet, complete with an assortment of various sign choices in the lattice vectors corresponding to choices of edge-dashing, and together with a precise correspondence of the two equivalence classes of these sign-choices.

We have hereby exhibited the correspondences:
\begin{equation}
 \vC{\begin{picture}(140,25)(-2,-2)
      \put(-13,10.5){$E_8$}
      \put(19,0){$\L_{E_8}^+$-basis}
      \put(19,20){$\L_{E_8}^-$-basis}
      \put(64,10.5){$e_8$}
      \put(27.5,10.5){$\l$}
      \put(107,10.5){$\Tw{\bs\l}$}
      \put(95,0){$({\bf A},{\bf B},{\bf A}_{\sss\ha\hb},{\bf B}_{\sss\ha\hb}|
                   \bs{\J}\1_{\sss\hK\ha})$}
      \put(95,20){$(\Tw{\bf A}_{\sss\hK\,\ha}|
                     \Tw{\bs\J}{}^+,\Tw{\bs\J}{}^+_{\sss\ha\hb},
                      \Tw{\bs\J}{}^-,\Tw{\bs\J}{}^-_{\sss\ha\hb})$}
      \put(-8,0){\includegraphics[height=23mm]{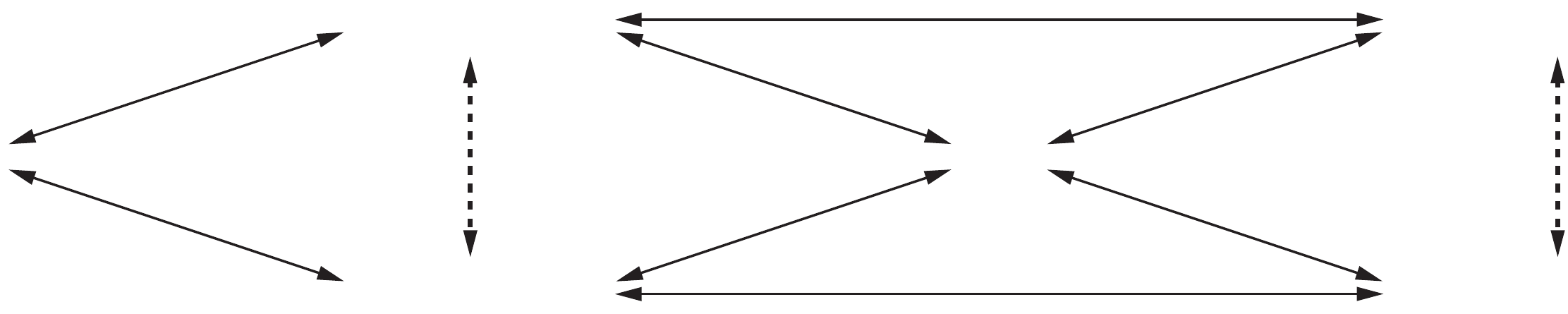}}
     \end{picture}}
 \label{eE8D}
\end{equation}
where the dashed arrows denote twist-isomorphisms: $\l$ between lattices, $\Tw{\bs\l}$ between superfields. Again, it would be interesting to see if the presented reconstruction of lattice bases may in fact be extended to a basis-independent reconstruction of abstract lattices. This would provide a direct relationship between the abstract $E_8$ lattices and the ultra-multiplet Adinkras, supermultiplets and the superfield Adinkras on the far right. Such a comprehensive and rigorous study is however well beyond our present scope.

The difference between an ultra-multiplet and its twisted variant is virtually identical to that between the chiral and twisted-chiral supermultiplets of worldsheet $(2,2)$-supersymmetry\cite{rGHR}. In particular, the isomorphism $\tw\l$ represents the fact that every model constructed with only ultra-multiplets can equally well be recast in terms of twisted ultra-multiplets. However, it may well be possible to construct a model that mixes both supermultiplets in an indecomposable way.
 The difference between chiral and twisted chiral supermultiplets was indeed usefully employable in this sense and led to a realization of ``almost product geometries'' in the target space\cite{rGHR}. We should thus like to conjecture that|unlike the purely quadratic Lagrangians\eqs{UM4}{UM6} and\eqs{ultraFLUX}{ultraPair}|the most general, fully interactive nonlinear and/or gauged $\s$-models do mix the ultra-multiplets and twisted ultra-multiplets in a nontrivial, indecomposable and ultimately useful way.

\vfill
\bibliographystyle{elsart-numX}
\bibliography{Refs}
\end{document}